\documentclass[fleqn,usenatbib]{mnras}

\usepackage{newtxtext,newtxmath}
\usepackage{subcaption}
\usepackage[T1]{fontenc}
\usepackage{ae,aecompl}

\usepackage{graphicx}	
\usepackage{amsmath}	

\usepackage{hyperref}

\title{3D Hydrodynamic Simulations of Large Scale Precessing Jets: Radio Morphology}

\author[M. A. Horton et al.]{
Maya A. Horton,$^{1}$\thanks{E-mail: mh17adw@herts.ac.uk}
Martin G. H. Krause,$^{1}$
and Martin J. Hardcastle$^{1}$
\\
$^{1}$Centre for Astrophysics Research, School of Physics, Astronomy and Mathematics, University of Hertfordshire, College Lane, Hatfield, AL10 9AB, UK\\
}

\date{Accepted XXX. Received YYY; in original form ZZZ}

\pubyear{2020}

\begin{document}
\label{firstpage}
\pagerange{\pageref{firstpage}--\pageref{lastpage}}
\maketitle

\begin{abstract}
The prospect of relativistic jets exhibiting complex morphologies as a consequence of geodetic precession has long been hypothesised. We have carried out a 3D hydrodynamics simulation study varying the precession cone angle, jet injection speed and number of turns per simulation time. Using proxies for the radio emission we project the sources with different inclinations to the line of sight to the observer. We find that a number of different precession combinations result in characteristic `X' shaped sources which are frequently observed in radio data, and some precessing jet morphologies may mimic the morphological signatures of restarting radio sources. We look at jets ranging in scale from tens to hundreds of kiloparsecs and develop tools for identifying known precession indicators of  point symmetry, curvature and jet misalignment from the lobe axis and show that, based on our simulation sample of precessing and non-precessing jets, a radio source that displays any of these indicators has a 98\% chance of being a precessing source. 
\end{abstract}




\section{Introduction}
\label{sec:intro}

Systems of binary supermassive black holes are believed to be a natural consequence of galactic evolution \citep*[e.g.,][]{begelman80,mayer17,tremmel18}, where dynamical friction in the wake of major mergers results in the greatest gravitational masses being slowed down by gas, dust and stars until they settle in an orbit \citep{chandra03}. 

In the case of binary systems in active galaxies, it is thought that geodetic precession of the black-hole spins will cause the jets to re-orient periodically on a timescale that depends on the orbital separation \citep{begelman80}. Such morphological structures have been studied with a ballistic jet model \citep{gower82, horton20a}, and similar structures have potentially been observed in kiloparsec-scale radio jets \citep{krause18}. 

\cite{krause18} identified four signatures of jet precession: 1) S-shaped, or radial, symmetry between jet and counterjet (S); 2) Jet curvature (C); 3) Jet at edge of lobe (E), (e.g., misalignment between jet and lobe axis) ; and 4) Multiple, or wide, terminal hotspots (H), and explored the incidents of these signatures. However, \cite{horton20a} modelled Cygnus A with a ballistic model and found that the structure of the jet and lobes could not be explained by ballistics alone. In real-world radio sources, hydrodynamic processes in lobes can push jets away from their ballistic paths, and may also disrupt them. Still, hydrodynamic models have also shown the development of characteristic precession signatures \citep[e.g.,][]{cox91,donohoe16, smith19}, whilst such signatures are absent in simulations of non-precessing jets \citep*[e.g.,][]{english16}.

Here we show 3D hydrodynamic simulations and synthetic radio maps of precessing jets with parameters similar to ones suggested by the precession interpretation of 100 kpc-scale radio sources \citep{krause18}. This would correspond to parsec-scale orbital separations, if the precession was caused by geodetic spin precession in binary supermassive black-hole systems. 

\section{Simulations}\label{sec:sims}
\subsection{Hydrodynamic setup}\label{subsec:pluto}
\begin{figure}
    \centering
    \includegraphics[width=\columnwidth]{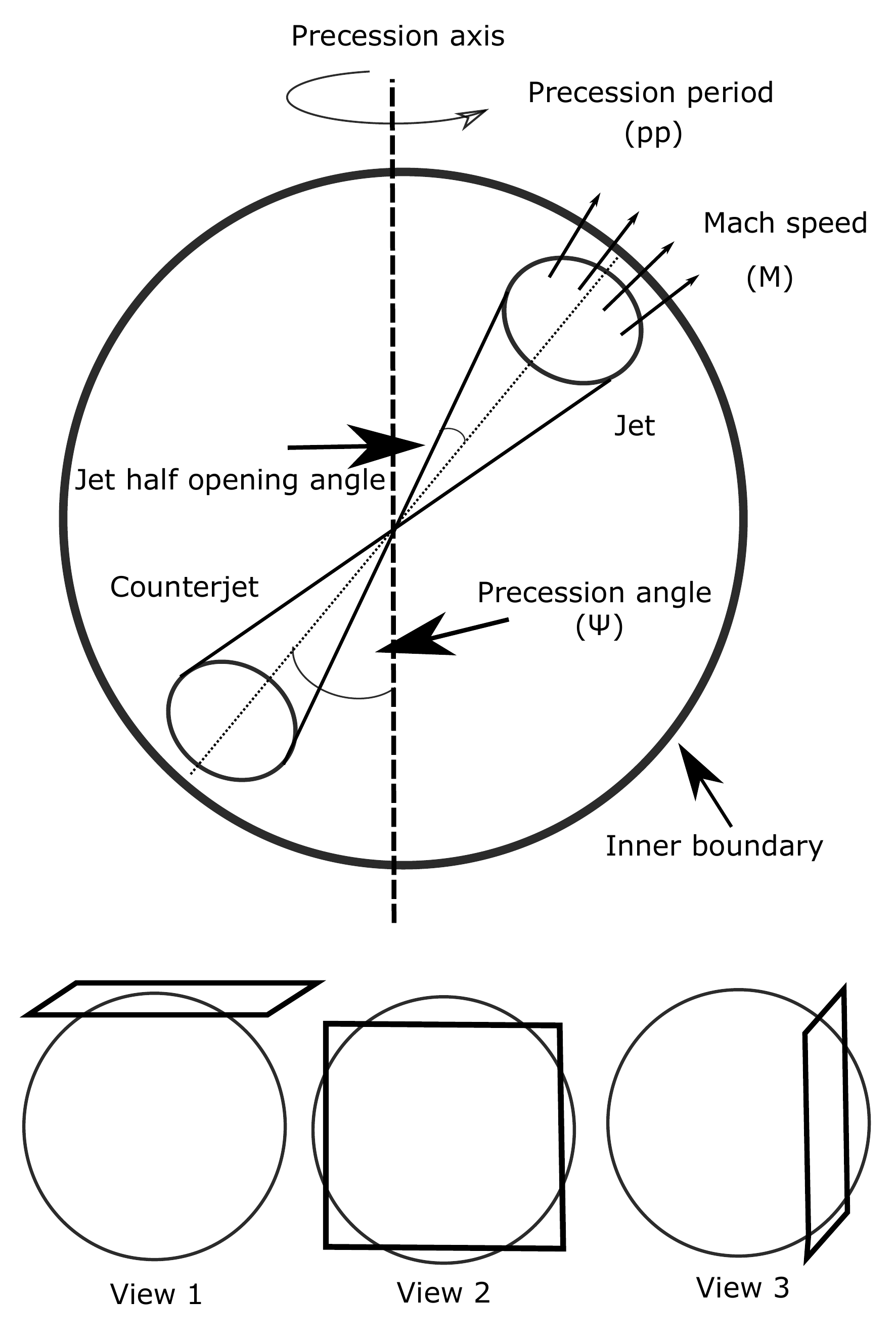}
    \caption{Schematic showing spherical simulation setup consisting of a jet and counterjet being injected along precession cone opening angle $\psi$ with a precession period $pp$ and Mach injection speed $M$. The three spheres represent the three projected views chosen for analysis throughout this paper: View 1 (top-down), View 2 (face-on), and View 3 (side on).}
    \label{fig:jet_diagram}
\end{figure}

We used the freely available PLUTO\footnote{\href{http://plutocode.ph.unito.it}{http://plutocode.ph.unito.it}} hydrodynamic code \citep{mignone07}, version 4.3, running the HD physics module with a two-shock \verb hllc  Riemann solver. Time-stepping uses $2^\mathrm{nd}$-order Runge Kutta (RK2) with a Courant-Friedrichs-Lewy (CFL) number of 0.2. Adaptive Mesh Refinement (AMR) was not used. The main parameter study was set up on a 512 x 256 x 512 grid using spherical polar coordinates (using the physics convention of $r, \theta, \phi$ corresponding to radial, polar and azimuthal angles) which were later reprojected into Cartesian coordinates. Conical jets were injected into a uniform density environment; the intersection of the cones with the inner boundary (see grid setup below) of the computational volume means that the jets appear at two oppositely placed spots on the inner boundary of the spherical co-ordinate system, which rotate about the precession axis (see Subsection~\ref{subsec:params} for details) at a rate determined by the precession period, where $pp = 1$ corresponds to the simulation time. All jets had a jet cone half opening angle of 5$^\circ$, as appropriate for Fanaroff-Riley II radio jets \citep{krause12}, and a Mach number $M \gg 1$ that corresponds to the speed at which the jet was injected (see Fig.~\ref{fig:jet_diagram}) and were injected with the same initial density and pressure as the central values for the computational volume (see below); the jets then naturally recollimate downstream. A counterjet was injected on the opposite side of the grid with slight time-varying perturbations to break up the symmetry. 

The grid extent in code units was set such that $0.2 \le r \le 5$, $0
\le \theta \le \pi$ and $0 \le \phi \le 2\pi$. The boundary conditions
were set to reflective (inner $r$, other than where the jet is being
injected) and outflow (outer $r$), periodic at the $\phi$ boundaries
and reflective with inversion symmetry for the velocity on the axis
($\theta =0,\pi$ ).

The computational volume is initialized in simulation units to a
density $\rho$ of 1 and a pressure $p$ of $1/\gamma$, where $\gamma=5/3$,
and the simulation unit of speed is the sound speed in this medium.
The uniform-density approximation here does not correspond to the
environments of real radio sources, but nevertheless we can estimate a
rough scaling of our simulation units to physical units. Let us assume
that the unit density $\rho$ corresponds to $n_p = 10^3$ protons m$^{-3}$ and the
temperature of the gas is $T = 10^7$ K (comparable to the temperature of
the hot gas in a group of galaxies, which radio galaxies often
inhabit). For a helium-hydrogen plasma the pressure corresponding to 1 simulation unit is then $\frac{5}{3} \times 2.3 n_p kT$, where the factor 2.3 gives the total number of particles per proton: this gives $p
= 5.3 \times 10^{-13}$ Pa, and an environmental pressure of $p/\gamma = 3.2 \times 10^{-13}$ Pa. We are free to choose the physical scale.
If we set the outer radius of the simulations to 300~kpc, as adopted
by \cite{hardcastle13}, which corresponds to a large,
well-resolved radio galaxy, then our simulation unit of distance is 60~kpc. The simulation sound speed is $v_s = \sqrt{\gamma kT/m}$ where
$m$ is the mean mass per particle, $m = 0.61m_p$, giving $v_s = 480$~km
s$^{-1}$. The scale choice means that one simulation time unit is $t = 60 \times
3.1 \times 10^{19} / 480 \times 10^3 = 3.9 \times 10^{15}$ s $=1.2
\times 10^8$ years. We use code units throughout the paper, as this
conversion to physical units is only one possible choice and is
necessarily somewhat arbitrary.

In these units the jet is injected with the ambient density and pressure at a radius of 12 kpc and at that point has a radius of 1 kpc. The kinetic power of the $M=100$ jet in physical units is then $3 \times 10^{38}$ W, a reasonable power for a Fanaroff-Riley class II object (e.g. \citealt{hardcastle13}). The power of the $M=50$ jets would be a factor 8 lower.

We ran all simulations to an initial simulation time of 0.3 (simulation units), writing out the PLUTO output every 0.001 simulation time units. Five additional simulations were chosen for their slower growth rates and extended out to a simulation time of 0.6 For convenience we refer to this time unit as the timestep in what follows, so the primary simulation duration is 300 timesteps whilst the longer runs are 600; for the example scaling to physical units given above, one timestep is $1.2 \times 10^5$ years. For each simulation group we ran an additional ``straight'' jet simulation with the same injection angle and Mach speed, but with no precession. These have been used as controls at every stage of analysis. 


All runs were performed on the University of Hertfordshire High Performance Computing cluster \footnote{\href{https://uhhpc.herts.ac.uk}{https://uhhpc.herts.ac.uk}}.

\subsection{Parameter study}\label{subsec:params}
 
\begin{table*} 
\caption{List of simulations run during the parameter study.}
\label{table:param}
\centering
\begin{tabular}{l r r r r r }
\hline
Simulation name & Cone angle & Mach number & Precession period  & Simulation time \\
& (degrees) & & (simulation units) & (simulation units) & \\
\hline
15\_50\_STR 	&	15	&	50	&	$\infty$	& 0.3  \\	
15\_50\_1 		&	15	&	50	&	1.0	 		& 0.3  \\
15\_50\_02   	&	15	&	50	&	0.2	 		& 0.3  \\
15\_50\_02\_L  	&	15	&	50	&	0.2	 		& 0.6  \\
15\_100\_STR	&	15	&	100	&	$\infty$	& 0.3  \\ 	
15\_100\_1  	&	15	&	100	&	1.0	 		& 0.3  \\
15\_100\_02  	&	15	&	100	&	0.2	 		& 0.3  \\
30\_50\_STR 	&	30	&	50	&	$\infty$ 	& 0.3  \\
30\_50\_1 		&	30	&	50	&	1.0	 		& 0.3  \\
30\_50\_02 		&	30	&	50	&	0.2	 		& 0.3  \\
30\_50\_02\_L 	&	30	&	50	&	0.2	 		& 0.6  \\
30\_100\_STR	&	30	&	100	&	$\infty$	& 0.3  \\
30\_100\_1    	&	30	&	100	&	1.0	 		& 0.3  \\
30\_100\_02   	&	30	&	100	&	0.2	 		& 0.3  \\
30\_100\_02\_L  &	30	&	100	&	0.2	 		& 0.6  \\
45\_50\_STR 	&	45	&	50	&	$\infty$	& 0.3  \\
45\_50\_1 		&	45	&	50	&	1.0	 		& 0.3  \\
45\_50\_02 		&	45	&	50	&	0.2	 		& 0.3  \\
45\_50\_02\_L 	&	45	&	50	&	0.2	 		& 0.6  \\
45\_100\_STR	&	45	&	100	&	$\infty$	& 0.3  \\
45\_100\_1 		&	45	&	100	&	1.0	 		& 0.3  \\ 
45\_100\_02 	&	45	&	100	&	0.2	 		& 0.3  \\
45\_100\_02\_L 	&	45	&	100	&	0.2	 		& 0.6  \\
\hline
\end{tabular}
\end{table*}

The parameters we varied are $pp$, the precession period (1 and 5 turns per 300 timesteps), $M$, the jet injection speed (Mach 50 and Mach 100), and $\psi$, the precession cone opening angle (15$^\circ$, 30$^\circ$ and 45$^\circ$).  These parameters were chosen because we expected them to produce structures corresponding to a range of morphologies where precession indicators may be present (see Table~\ref{table:param}). We expected the parameter choices to range from little or no signature to those with highly complex structures and multiple indicators of precession. Early stages of simulations, when source age is small compared to precession period, may be scaled to larger radio sources with slower precession rates. Also, the symmetric lobe structures observed for some precession candidates in \cite{krause18} suggest precession for many turns. In addition, we ran a subsample of jets for twice the length (0.6) simulation time. These are denoted with \_L where applicable and were chosen because of their low growth rate along the $r$ direction, in order to assess the consequences of isotropism on precession indicators. 

\subsection{Synthetic radio maps}\label{subsec:data}
To make the synchrotron visualisation (shown in blue in the movie images), we converted pressure to emissivity following the method of \cite{hardcastle13}. We take the synchrotron emissivity as being proportional to $p^{1.8}$ and integrate along the chosen line of sight to obtain the radio map. This works well for the radio lobes, but not for the jets. The reason is that jets are strongly affected by relativistic beaming and may have higher magnetic field strengths than the radio lobes. Their visibility depends also on particle acceleration processes in the jet, which are not understood in detail yet \citep[e.g.,][]{hardcastle16,sun18}. To visualise the jets we therefore took any structure with a Mach number greater than half of the jet injection speed. Once calculated, we projected the 3D spherical grid into a 3D Cartesian grid and integrated the emission along the line of sight for three viewing directions along the three Cartesian coordinate axes (see Fig.~\ref{fig:jet_diagram}). The jets, in purple, are shown as projections of the Mach number combined with the jet tracer. 

We used these synthetic 2D images in all subsequent analysis using a box size of $1024 \times 1024$ pixels for each view. We treated this as ideal and limited only via numerical resolution rather than including the beam size of a realistic radio image. Since our simulations are in polar coordinates the resolution is radius-dependent,  but we do not expect this to affect our analysis. The dynamic range is $10^5$ for both Mach 50 and 100 jets. Since radio images with e.g. the VLA routinely achieve a dynamic range of $10^4$, all main features should be observable; future instrumentation, such as the SKA, is likely to achieve dynamic ranges of $10^6$ or more \footnote{\url{https://astronomers.skatelescope.org/documents/}}. 

We chose three views, corresponding to three adjacent faces of a cube: these are labelled \textit{top down} or View 1, \textit{face on}, or View 2, and \textit{edge on} (View 3). Being top down, View 1 was always the most projected whilst View 3 often showed characteristics of straight jets for some portion of the source time. This reflects a correlation with the initial precession phase. 

Movies of these simulations are available at: \url{https://www.extragalactic.info/precessingjets}

\subsection{Determination of lobe physical structure}
Once the simulations have been reprojected into Cartesian space, we divide each of the simulated images for each timestep into 'north' and 'south' (top and bottom) and for each half we find the distance corresponding to the most distant point in the lobe, which we call the lobe length, and the total number of pixels in the lobe, which we call the lobe area; then the axial ratio is defined as the number of pixels divided by the lobe length squared (so it is low for long thin lobes and high for short wide ones). These numbers are tabulated for both lobes for each timestep.

We corrected for the variation in lobe position angle as a result of the varying precession angles and projection directions we used by finding a characteristic angle on the sky for the lobes in each view. In detail, we found the covariance matrix of the $x$ and $y$ co-ordinates of regions that appear inside the lobe, and then used the eigenvector of that matrix with the largest eigenvalue as the lobe direction. The images were rotated through the angle of this eigenvector on the sky before analysis. This is intended to mimic what would be done by observers, who would refer jet properties to a characteristic lobe axis.

\subsection{Definition of precession markers}
We looked at three of the four precession markers as indicated by \cite{krause18}: point symmetry between jet and counterjet (S); jet curvature (C); and lobe axis misalignment (E). Hotspot structure was not considered in this paper because we do not model particle acceleration, and so cannot visualise hotspots in a way that can be compared accurately with observations.

Lobe axis misalignment (E) was examined by looking at the proportion of time that the jet was at the edge of the lobe. At each point along the jet we took the absolute value of the distance between the lobe centre and the jet centre, normalizing by the width of the lobe. We then averaged this over the length of the jet for each timestep to obtain index (E). 

The curvature (C) is assessed by fitting a straight line to the entire jet path and calculating a curvature indicator of the form
\[
C = \frac{\chi^2}{N} = \frac{1}{N} \sum \left[y_i - (ax_i + b)\right]^2
\]
where $N$ is the total number of data points in the jet, the $y_i$ are the displacement from the jet lobe axis in pixels, $x_i$ are the radial distances in pixels, and $a$ and $b$ are the parameters of a straight line obtained by minimizing $\chi^2$. The higher the number, the worse the fit for the straight line. This was chosen simply because one jet could experience multiple forms of curvature throughout its lifecycle, and more prescriptive approaches would fail to take this into account.

Point (rotational) symmetry (S) was determined in the same way as lobe misalignment, but taking the signed distance between the lobe and jet centres. We consider only points where both a jet and counterjet are seen and we subtract the best-fitting straight line fit to the jet and counterjet as described above. Then for a given distance along the jet in pixels $x_i$, we have displacements $d_Ni$ and $d_Si$, and the product $d_Ni(-d_Si)$ is positive if the jet has S-symmetry. The curvature indicator is then given by
\[
S = \sqrt{\frac{\sum (d_Ni)(-d_Si)}{Nw^2}}
\]
where $N$ here is the number of points in the jet and $w$ is the mean width of the lobe in pixels, and we take a signed square root so that the presence of point symmetry produces a positive result, whilst negative values of the statistic (corresponding to mirror symmetry) give negative values of $S$. 

The values of these indicators are computed for each timestep in the simulation and for each of the three views. The fast straight jets, only, run off the grid before the end of the run time of 0.3 simulation time units we chose as the basis of our statistical analysis. The boundary is open and thus this effect reduces the lobe pressure somewhat which may contribute to the sideways expansion of the jet and affect the precession indicators. We have verified, however, that the statistics of the precession indicators for the straight jets depend only very weakly on simulation time. We therefore decided not to exclude times when the jet has run off the grid in the statistical analysis below, in order to have a uniform time base.

\subsection{Numerical resolution dependence}
Given that the simulations are in spherical coordinates, the cells closer to the centre of the grid have a higher resolution than those towards the edge. This has been shown to influence the realisation of hydrodynamic structures \citep{krause01}, and since the Mach 50 jets naturally do not grow as far along the grid as their faster counterparts, it is important to confirm that any differences and similarities between Mach 50 and 100 jets is resolution-independent. 

We picked a dynamically complex Mach 100 jet (45\_100\_1), which we used as a baseline: for both low (LR) and high (HR) resolution jets the radial component $r$ remained constant at 512, whilst $\phi$ and $\theta$ were decreased or increased by 1.5, or by 2 for the very high resolution (VHR) jet. We did the same for a non-precessing jet with the same characteristics. These are summarised in Table~\ref{table:res_grid}.



\begin{table} 
\centering
\begin{tabular}{l c c c}
\hline
Simulation name & \multicolumn{3}{c} {Grid Size}  \\
& $r $& $\theta$ & $\phi$ \\
\hline
45\_100\_STR\_LR 	&	512 & 192 & 384	\\ 
45\_100\_STR 		&	512 & 256 & 512	\\ 
45\_100\_STR\_HR 	&	512 & 384 & 768	\\
45\_100\_1\_VHR 	&	512 & 512 & 1024\\
45\_100\_1\_LR   	&	512 & 192 & 384	\\
45\_100\_1         	&	512 & 256 & 512	\\
45\_100\_1\_HR 		&	512 & 384 & 768	\\
45\_100\_1\_VHR 	&	512 & 512 & 1024\\
\hline
\end{tabular}
\caption{Grid sizes $r$, $\theta$ and $\phi$ for simulations in resolution study.}
\label{table:res_grid}
\end{table}

To assess the influence of resolution on lobe growth and jet structure, we repeated the analysis undertaken for the parameter study on all of these jets.

\section{Results}
\begin{figure*}
    \centering
        \includegraphics[height=8in]{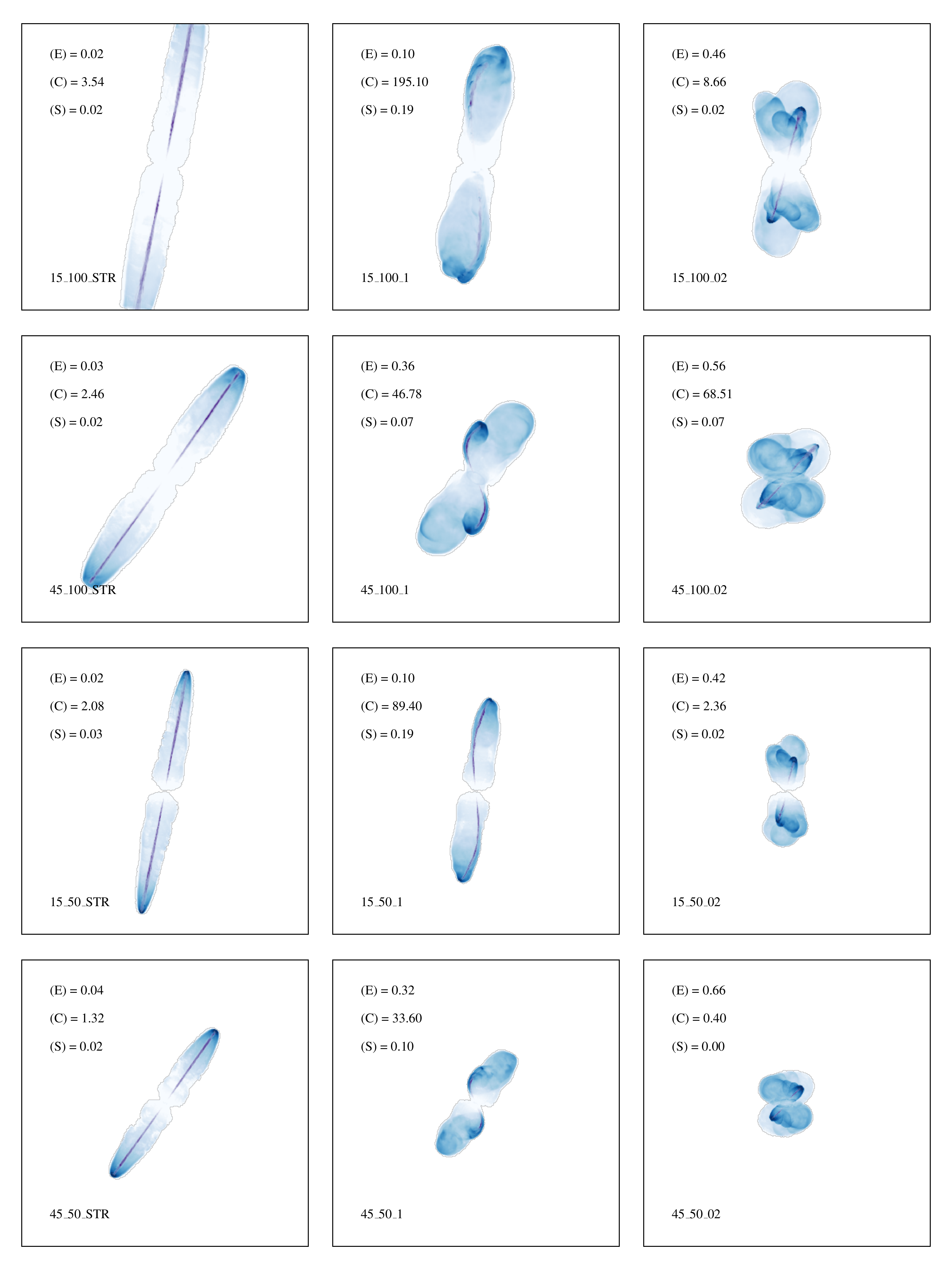}
    \caption{Synthetic radio maps from representative simulations, all at $ts = 200$ and the same fixed view (View 3). Linear purple features denote the jet paths as indicated by regions of high Mach number (see Section \ref{subsec:data}) for details. First column represents straight (non-precessing) jets; the second column shows jets with a single turn per 300 timesteps (e.g., have completed $2/3$ of a turn at this point) and the third shows those with five turns per 300 timesteps (and have completed 3.3 turns). The first and third rows correspond to a precession cone opening angle of 15$^\circ$ whilst the second and fourth open at 45$^\circ$. The first two rows are have a jet injection speed of Mach 100 whilst the remaining two are Mach 50. For each simulation we have given the indicator values for edgeness (E), curvature (C) and point symmetry (S), at that specific timestep. The dynamic range of the images is $10^5$, close to observable limits in real sources.}
    \label{fig:gridview}
\end{figure*}

\subsection{Morphology and jet structure}
Figure \ref{fig:gridview} shows a snapshot of 12 different simulation runs at the same timestep taken from 2/3 of the way through each simulation. Column~1 shows non-precessing jets; column 2 is for slowly precessing jets with a precession period equal to the source age, and column 3 shows jets with a precession period of 1/5 the source age. Rows~1 and 2 are for Mach 100 jets whilst rows 3 and 4 are for Mach 50 jets; therefore, rows 1 and 3 correspond to precession cone opening angles of 15$^\circ$ whilst rows 2 and 4 are at 45$^\circ$.  These result in a wide range of morphological structures. Many of these exhibit characteristics of real-world sources such as amorphous, restarting or X-shaped radio galaxies. Fast precessing objects can show straight jets in the shorter pair of X-shaped lobes whereas the more slowly precessing jets appear curved and bent by the lobe walls.

\subsection{Quantitative effects of precession and Mach number}
\subsubsection{Lobe growth and axial ratio}
\begin{figure*}
    \centering
    \includegraphics[width=\linewidth]{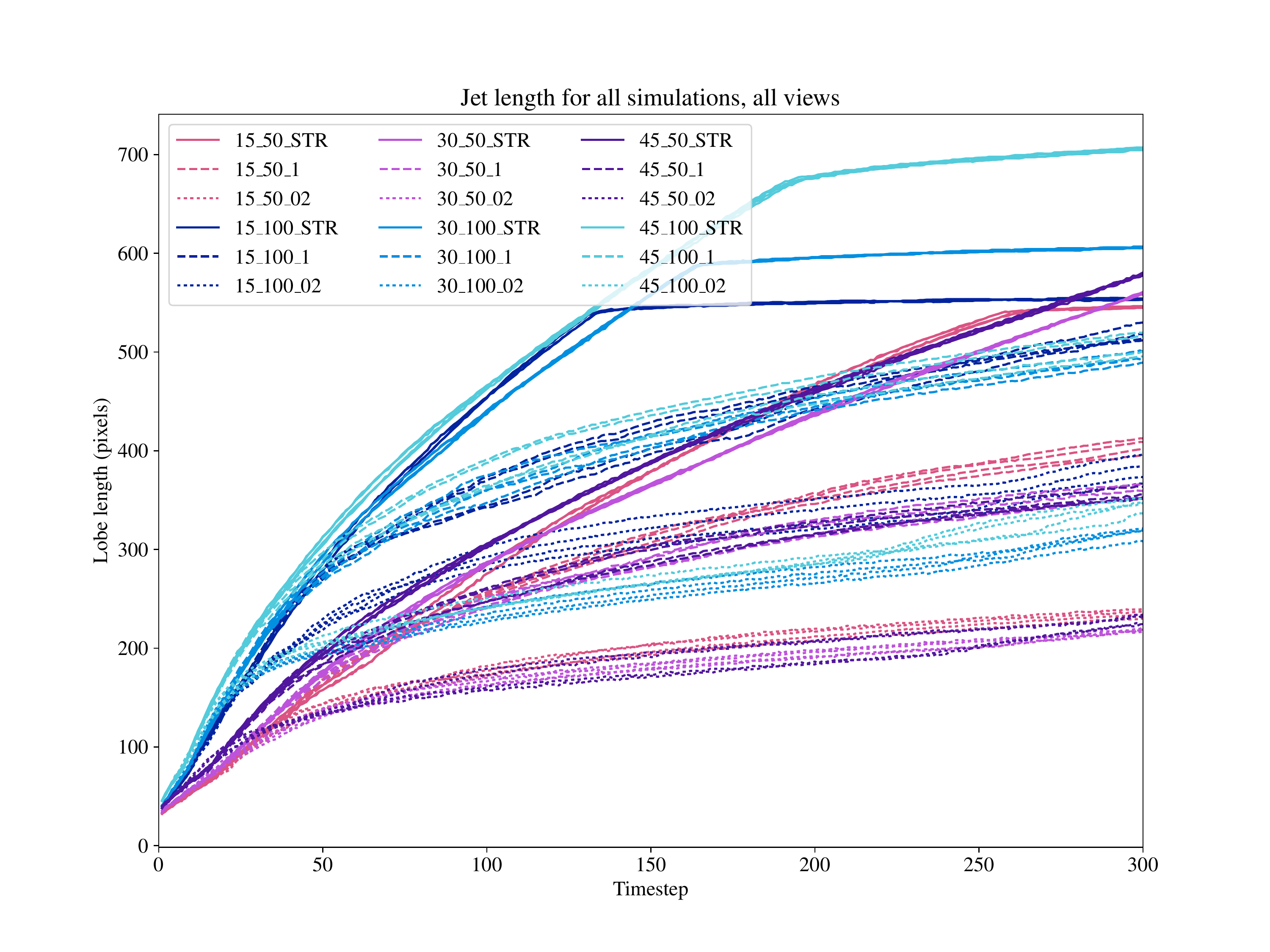}
    \caption{Lobe length growth for all simulations, corrected for projection effects. The flattened solid lines of the Mach 100 straight jets occur when the lobe expands beyond the edge of the grid. Each simulation shows both jet and counterjet for Views 2 and 3.}
    \label{fig:lobelength_allsims}
\end{figure*}

 Figure~\ref{fig:lobelength_allsims} shows numerous impacts on jet length as a consequence of precession cone opening angle and injection speed. Both jet and counterjet for Views 2 and 3 are shown. Jet injection angle varied in some simulations and was corrected for during analysis. Flattened lines indicate that a simulation has run off the grid, which only occurs for Mach 100 jets. Straight jets grow faster than precessing ones. 
 
 For all simulations, lobe growth rates decrease with decreasing precession period. Lower velocities also result in lower growth so the most compact sources are those with rapid precession, wide cone opening angles and slower Mach number. Given the complexity of parameter space there is often a trade-off between precession angle and velocity, since Mach 50 straight jets initially grow more slowly than slowly-precessing Mach 100 jets until approximately halfway through the simulation when the straight jets continue growing at almost the same rate and precessing jets flatten out. 
 
\begin{figure*}
    \centering
    \includegraphics[width=\linewidth]{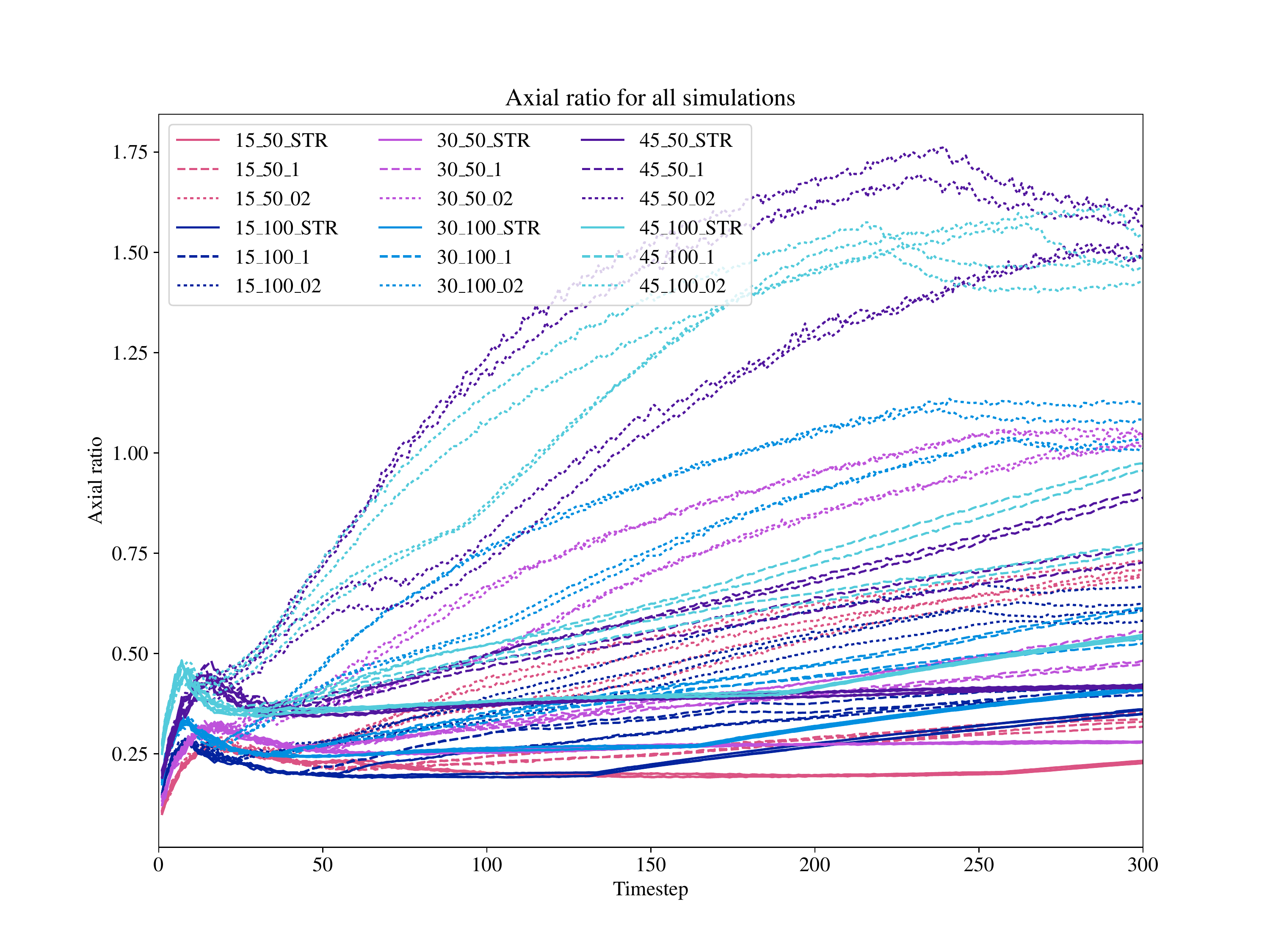}
    \caption{Changes in axial ratio for all simulations. Each simulation shows both jet and counterjet for Views 2 and 3.}
    \label{fig:lobegrowth_allsims}
\end{figure*}

 Axial ratio (Fig.~\ref{fig:lobegrowth_allsims}) shows similar trends. Assessment of axial ratio changes are initially unreliable for the first ~20 simulation time units of all simulations. This is a common feature of such simulations related to the necessarily unphysical initial conditions, which must relax before a realistic situation is established. After this, the straight jets remain at the lowest levels until they run off the grid. Jets with a precession period of 1 also have low axial ratios with some trends relating to Mach number. It is again important to note that both jet and counterjet, for two different projected views, are shown in this plot. Therefore some simulations -- for example, 45\_50\_02 -- show differing axial ratios depending on projection. These highlight the importance of remembering just how much apparent precession morphologies change depending upon their orientation along the line of sight. There is a clear, general trend for all axial ratios: radio lobes become significantly fatter, for faster precessing jets. 

\subsubsection{Precession indicators}
\begin{figure*}
    \centering
    \includegraphics[width=0.85\textwidth]{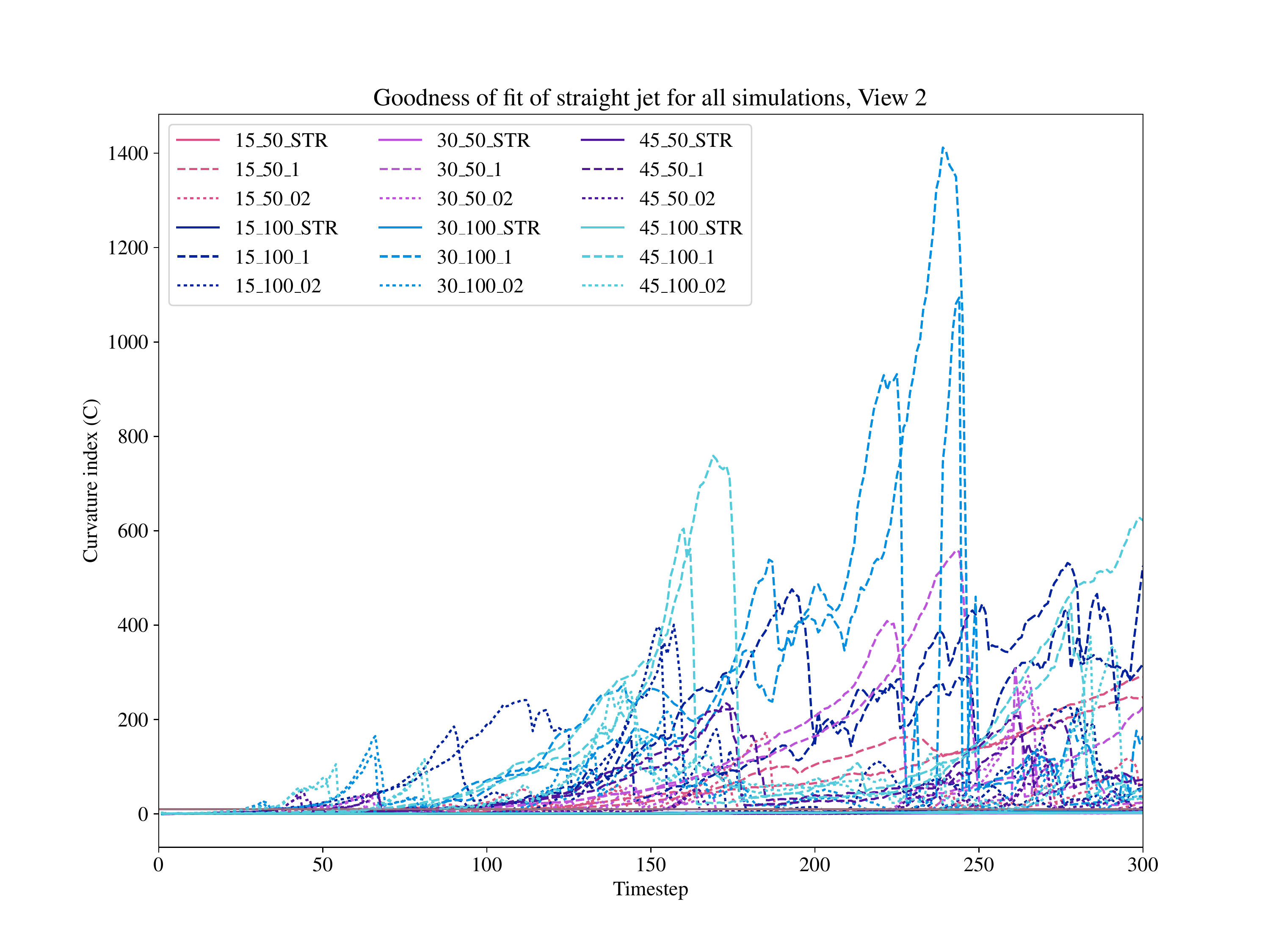}

    \includegraphics[width=0.85\textwidth]{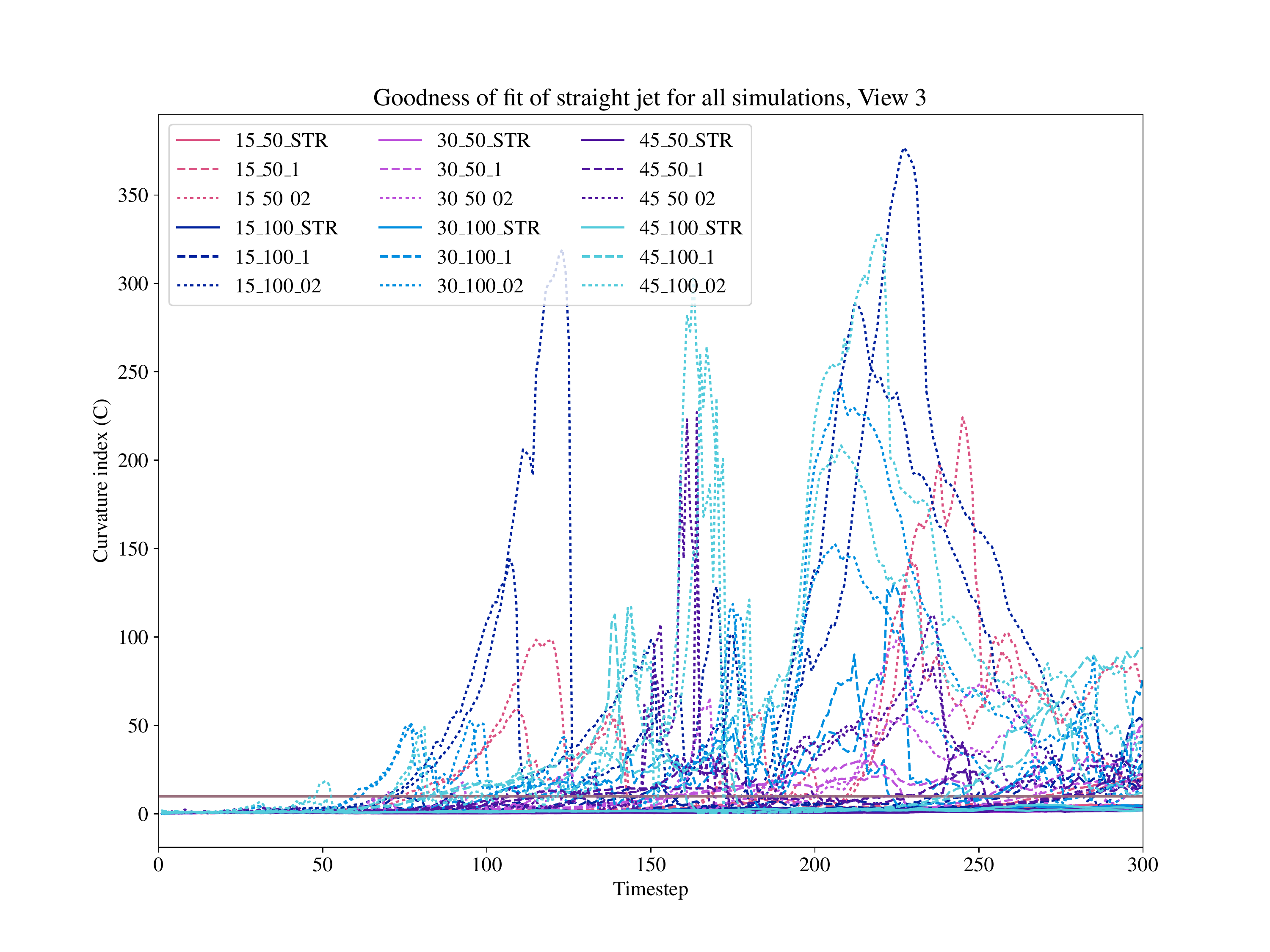}

    \caption{Curvature (C) precession markers for all simulations. Solid straight line shows indicator threshold of $C = 10$. Top panel shows `View 2' whilst bottom shows `View 3'.}
    \label{fig:allsims_curvature}
\end{figure*}

\begin{figure*}
    \centering
    \includegraphics[width=0.85\textwidth]{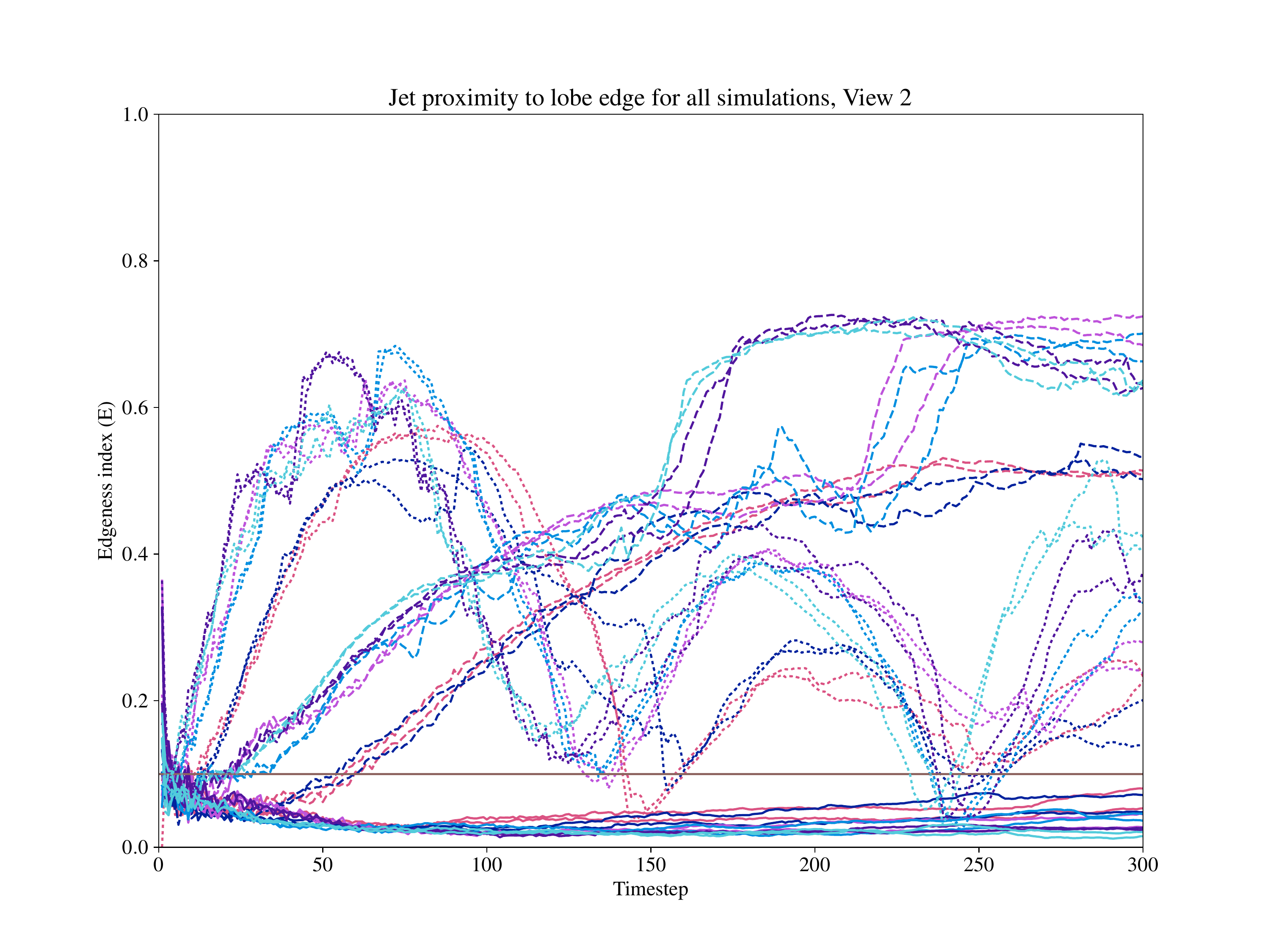}

    \includegraphics[width=0.85\textwidth]{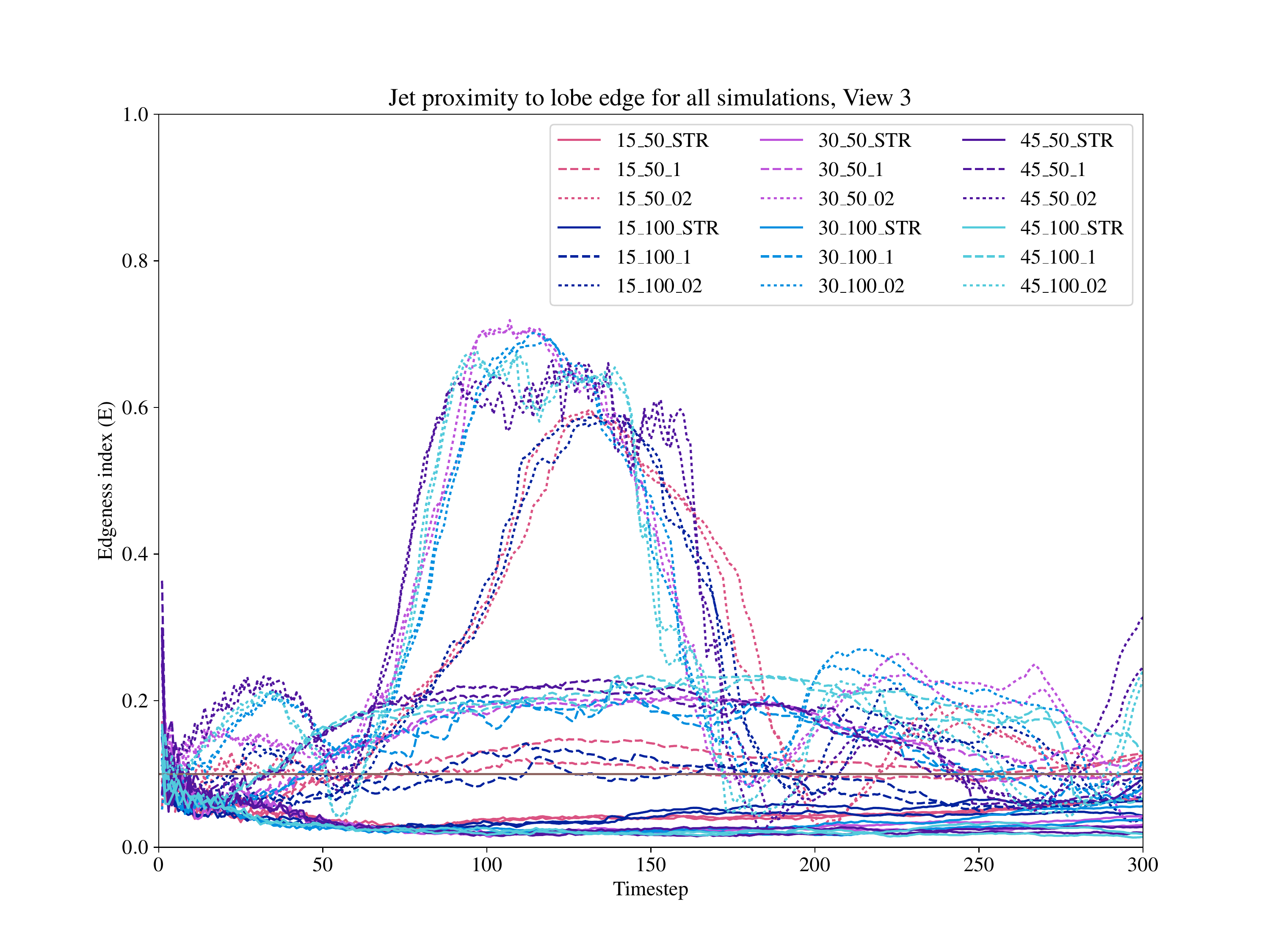}

    \caption{Mean lobe axis misalignment (E), expressed as position of jet throughout simulation relative to centre and edge of lobe where 0 is no curvature (the jet is in the centre of the lobe) and 1 is the jet is exactly at the edge of the lobe. This represents the variation throughout a single fixed side-on view for all simulations as described in Table~\ref{table:param}. Solid straight line shows indicator threshold of $E = 0.1$. Double lines of the same type indicate jet and counterjet. Top panel shows `View 2' whilst bottom shows `View 3'.}
    \label{fig:allsims_edgeness}
\end{figure*}

\begin{figure*}
    \centering
    \includegraphics[width=0.85\textwidth]{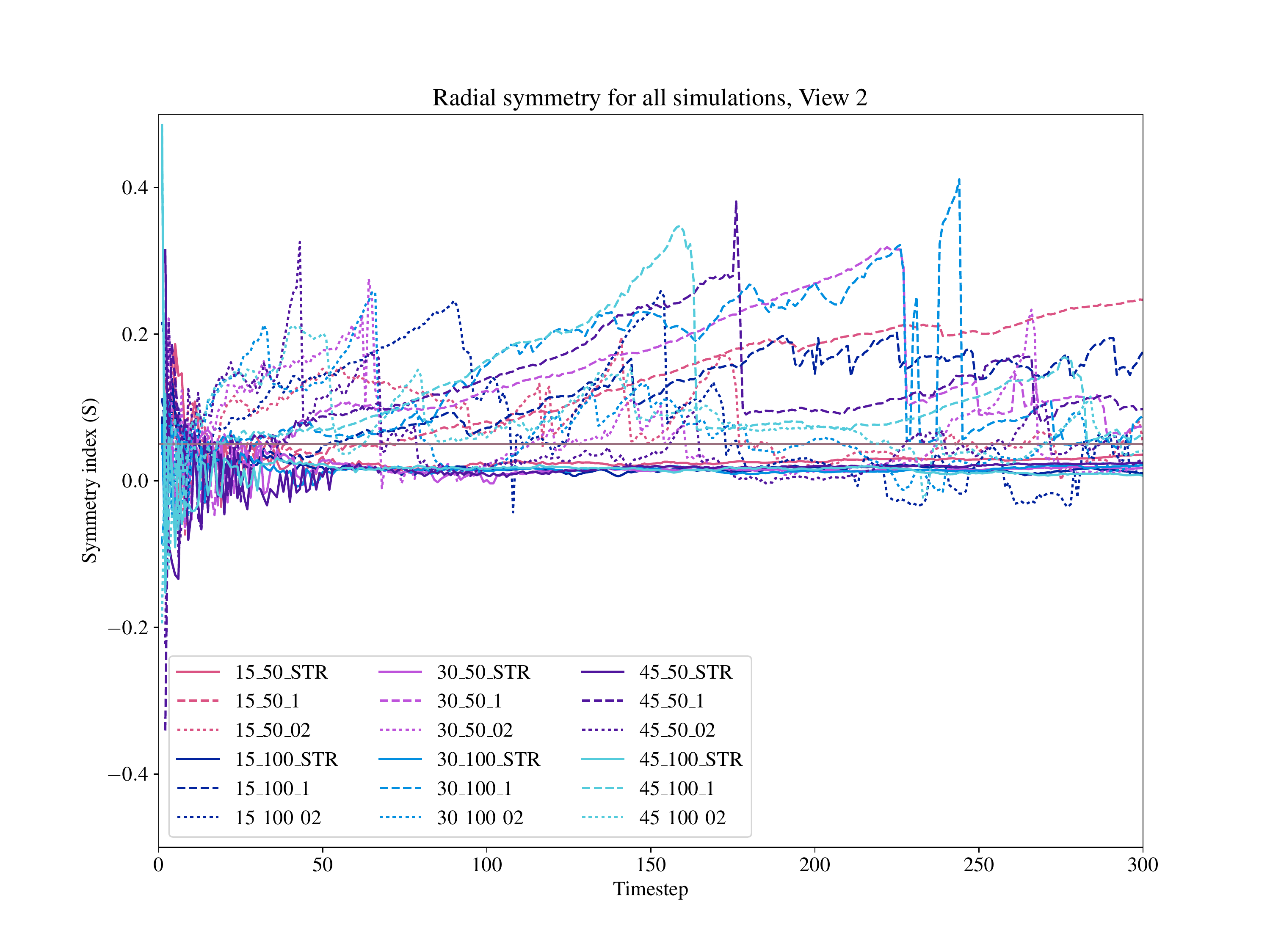}

    \includegraphics[width=0.85\textwidth]{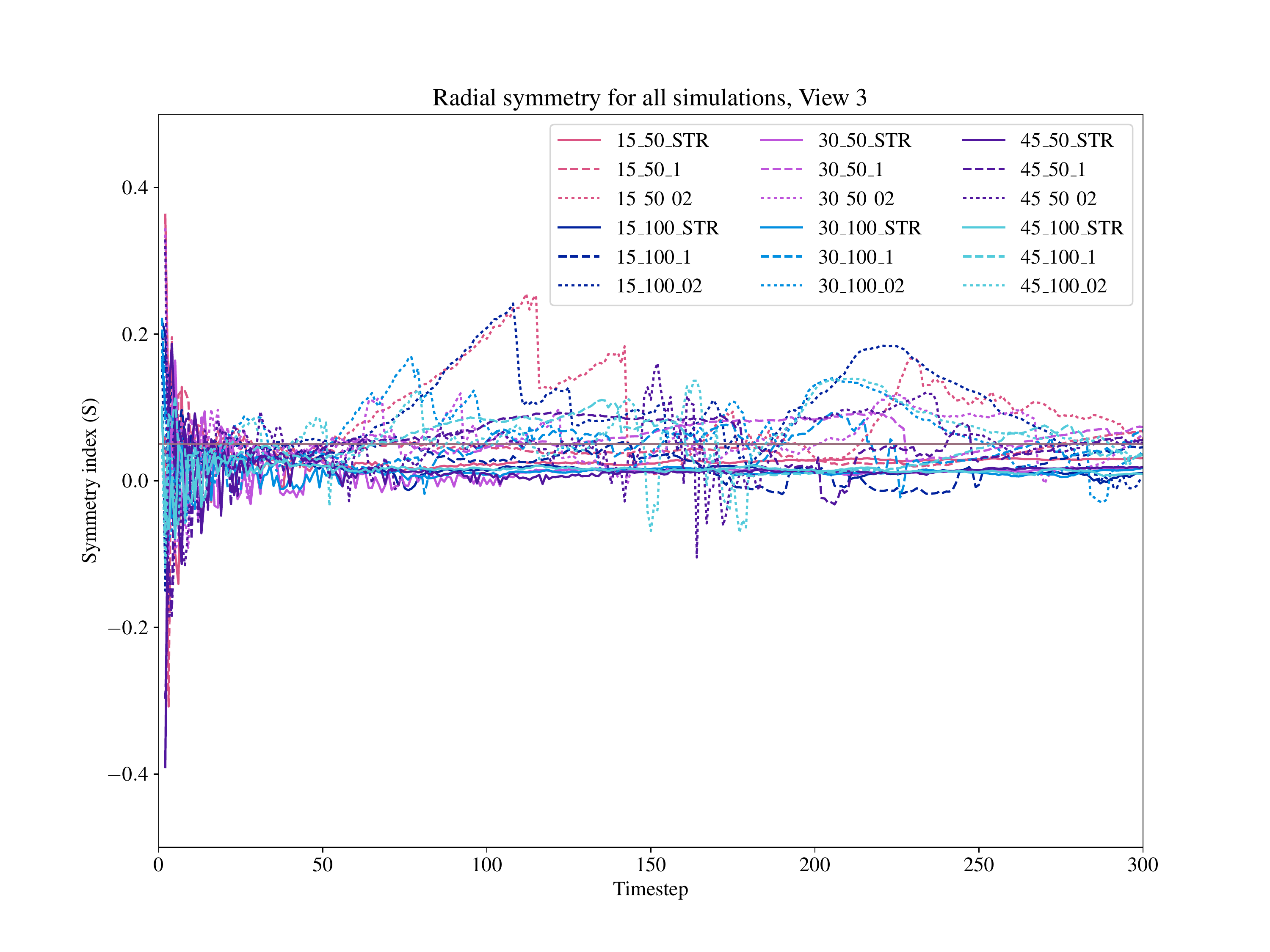}
    
    \caption{'S'-shaped symmetry (S) for `View 2' (top) and `View 3' (bottom) of all simulations. Solid straight line shows indicator threshold of $S = 0.05$. Legend as in previous plots.}
    \label{fig:allsims_sness}
\end{figure*}

Figures \ref{fig:allsims_curvature} to \ref{fig:allsims_sness} show the prevalence of precession indicators. Each figure shows all simulations for Views 2 and 3. View 1 has been omitted for space, and because the top-down view is highly projected for all simulations and hence less relevant for radio galaxies. Table~\ref{table:curve_frac} shows the percentage of time that each precession indicator exceeds a given threshold level for each jet, for all three views. We chose the thresholds to be $E = 0.1$, $C = 10$ and $S = 0.05$. These were chosen manually to reflect the extent which could not normally be reached by non-precessing hydrodynamics alone (except during very early periods of lobe expansion), and as such can be thought of as minimum values for precession. Higher threshold values would not have impacted straight jet detection and would have naturally increased the percentage of false negatives. Instead we chose to set values which provide us with the maximum amount of time a precession indicator could be detected from a precessing jet. These values also make sense intuitively. This can be seen from Fig. 2, where we give all precession indicator value for each simulation snapshot shown (compare also Fig. 9). For example, the fast precessing run 15\_100\_02 shows a jet that appears essentially straight for the given snapshot with a curvature indicator C=9, just below the threshold value C=10. Run 45\_50\_1 has C=34 and shows a clearly curved jet. The straight jets show a simple lobe-jet structure with E up to 0.04 in the snapshots shown. 15\_50\_1 and 15\_100\_1 both have E=0.1, i.e., are at the threshold value and clearly have the jet towards the edge of the lobe. For S, the shown straight jets have values up to 0.03, with no visual indication of S-symmetry, whereas run 45\_100\_1 clearly shows the symmetry and has S=0.07, supporting the choice of the threshold value at 0.05.

\begin{figure*}
    \centering
        \includegraphics[height=2.2in]{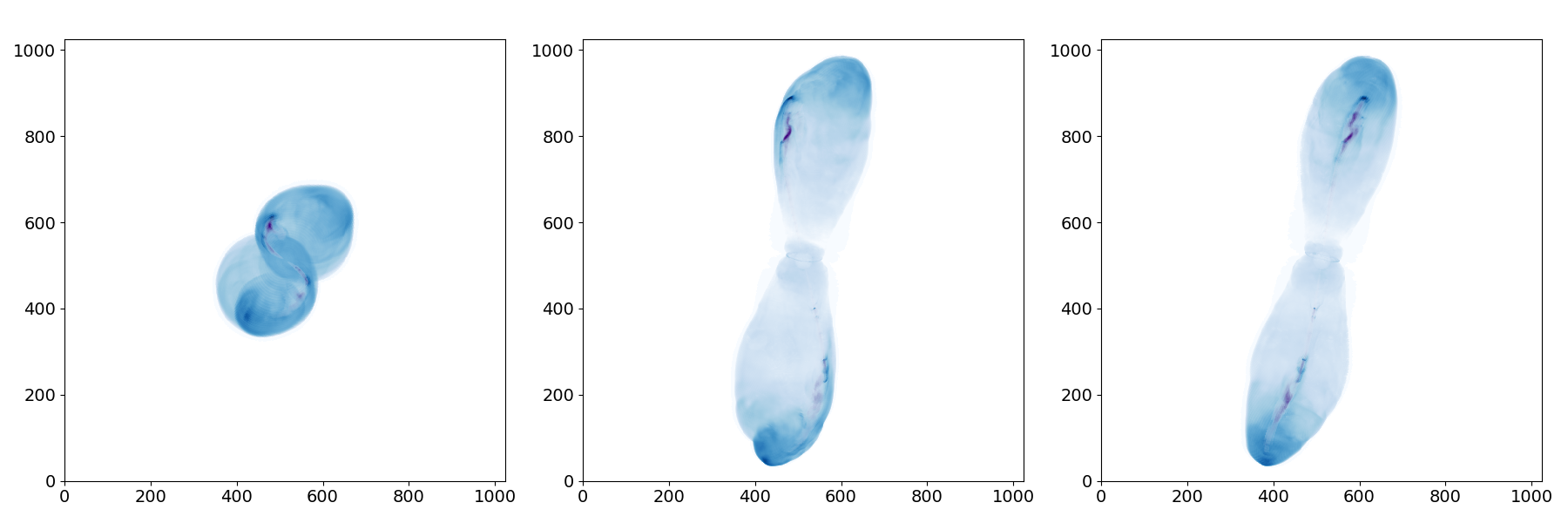}
    \caption{Three projected views from a $M = 100$, $\psi = 15^\circ$, $pp = 1$ jet close to the end of its simulation. The first panel shows a top-down view, and the second and third from two adjacent sides. The jet is purple whilst the blue indicates modelled synchrotron emission. By this point the jet is knotty and discontinuous.}
    \label{fig:jc_151001}
\end{figure*}

Jet curvature (C) (Fig.~\ref{fig:allsims_curvature}) is strongly dependent on view, with those jets with slower precession and higher jet speed (e.g., run 15\_100\_1) revealing little curvature for View 3 for most of the simulation (9\%; as opposed to 61\% curvature in View 2). This can be confirmed visually in Fig.~\ref{fig:jc_151001}, far panel, which shows a snapshot from that simulation highlighting that View 3 appears straight to a casual observer whilst top-down and edge-on views of the same source at the same time show strong curvature. The differences between View 2 and View 3 are much less pronounced for the fast precessing jets. This shows that the initial orientation is still important for the final morphology throughout much of the first turn. 

Even though there is this difference in viewing angle for the slowly precessing jets, it is striking that the curvature indices of the fast-precessing jets never get anywhere near the slowly precessing ones \ref{fig:allsims_curvature}. The maximum $C$ a fast-precessing jet ever reaches in any View (2 or 3) in our simulations is below 400, whereas the slowly precessing jets reach up to > 1400. This corresponds to the impression from Fig.~\ref{fig:gridview}, where the fast-precessing jets appear almost straight, whereas the slowly-precessing ones appear highly curved. The reason seems to be that the fast-precessing jets break up quickly whereas the slowly precessing ones tend to get bent at the lobe boundaries.

Given the interdependence of the precession indicators, it should not be surprising that edgeness (E) shows similar patterns. For example the same simulation, 15\_100\_1,  high edgeness indicators for 85.5\% and 81.3\% for Views 1 and 2 respectively, but for View 3 this is is 29.2\% at the given threshold (it is worth noting that the threshold is set at a minimum value, and even slight increases give 0\% precession for that view). View 3 shows all precession indicators much more consistently for greater precession cone angles. Straight jets are still clearly distinguished in these measures: no false positive occurs for more than 5\% of the time, and this is usually only for one indicator and one view. Conversely, false negatives can indeed be at zero for a single indicator, but for a chosen view never fall lower than 29\% for at least one indicator. Whilst it is therefore important to recognise that there is a population of sources for which precession may be hard to detect, examining trends across all indicators is likely to provide more evidence that precession is indeed occurring.

Fig.~\ref{fig:allsims_sness} gives the same variation of S-symmetry (S) for Views 2 and 3: After the initial relaxation -- which is more pronounced for jets with greater cone opening angles -- straight jets often show non-zero s-symmetry values of under 5\%. This is because of minor disturbances along the lobe axis; however, this can be corrected for by removing the first ~50 timesteps after the initial lobe expansion stage. We have kept this in for completeness; the effect can be seen clearly at the start of Fig.~\ref{fig:allsims_sness} where hydrodynamic forces of early inflation create both radial symmetry and asymmetry for a short time. Some of the more strongly precessing jets show slight dips towards zero as the precessing jets change direction, but for the bulk of the simulation time there is positive radial symmetry between the jets and counterjets. Our slowly precessing jets generally show more pronounced S-symmetry, which may again be due to the fact the the fast precessing ones tend to break up and re-form. There is again a dependence on view for the slowly precessing jets, suggesting that again that precession indicators may not show consistently during the first turn. 

\begin{figure*}
    \centering
        \includegraphics[height=9in]{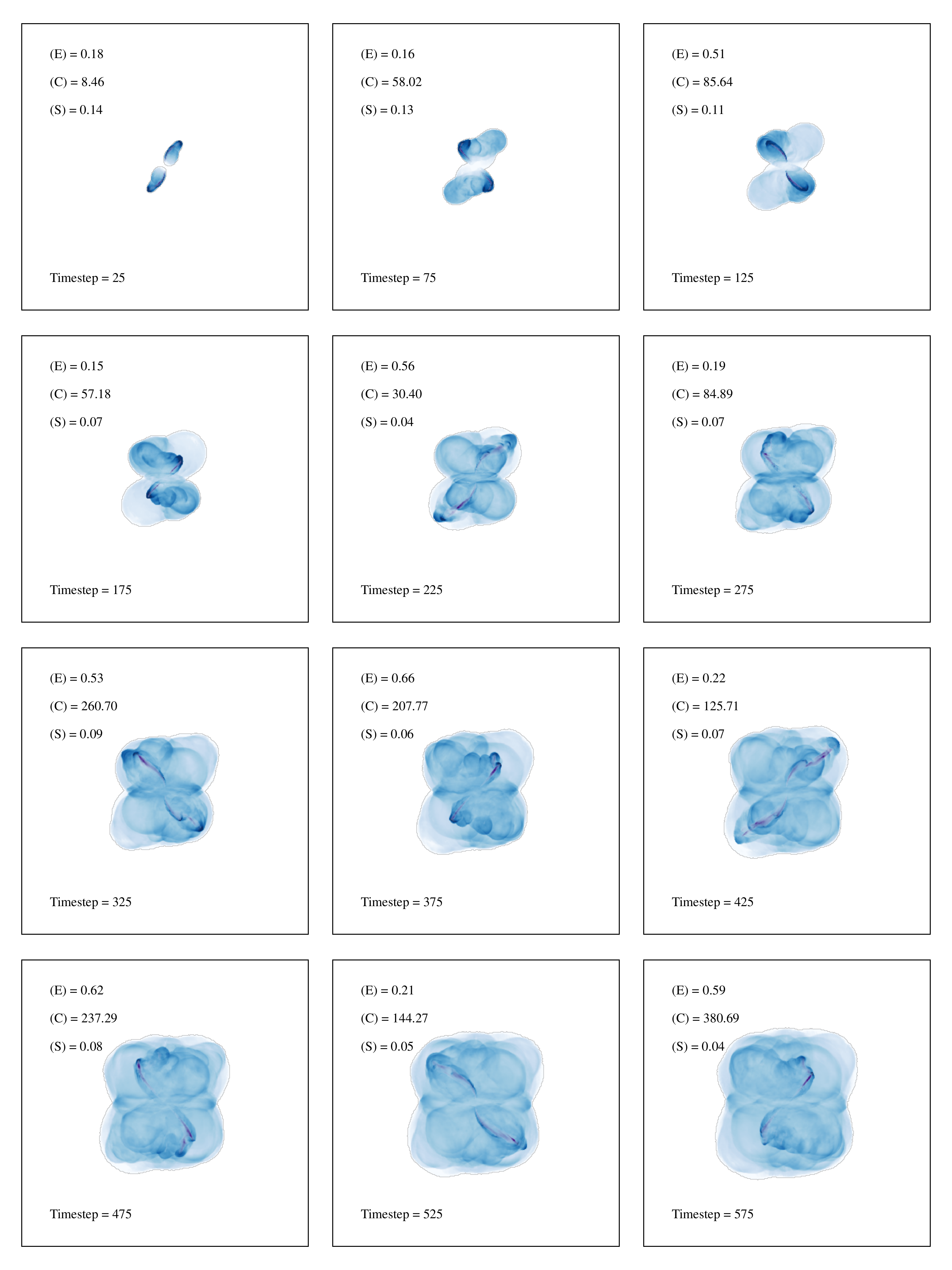}
    \caption{Timeseries evolution of a $M = 100$, $\psi = 45^\circ$, $pp = 02$ jet with a $0.6$ simulation time, as shown from View 2. Values of (E), (C) and (S) are shown for each snapshot.}
    \label{fig:gridview-45_100_02_L}
\end{figure*}

There is very little impact on indicator levels for the long runs (\_L). Since the jets are typically isotropic after one turn, doubling the simulation time typically makes little difference to the amount of time that a given precession indicator is observed. This is confirmed in Fig.~\ref{fig:gridview-45_100_02_L} where the (E), (C) and (S) values are often close to, or even beneath threshold values despite the obvious complexity of the lobe structure. 


\begin{table*} 
\caption{Percentage of simulation time where jet shows precession indicators that exceed the selected thresholds ($E = 0.1$, $C = 10$ and $S = 0.05$. Simulations indicated in brackets and marked with \_L were extended to twice the simulation time.)}
\label{table:curve_frac}
\centering
\begin{tabular}{lrrrrrrrrr}
\hline
Simulation name & \multicolumn{3}{c}{Jet at edge (E)} & \multicolumn{3}{c}{Jet curved (C)} &\multicolumn{3}{c}{Jet S-symmetric (S)}\\
& View 1 & View 2 & View 3 & View 1 & View 2 & View 3 & View 1 & View 2 & View 3   \\
\hline
15\_50\_STR& 0.3& 0.7& 0.2& 0.0& 0.0& 0.0& 1.0& 3.7& 4.0\\
15\_50\_1& 86.5& 81.2& 61.8& 56.9& 61.0& 9.0& 83.3& 83.6& 17.4\\
15\_50\_02& 91.2& 91.5& 79.8& 72.4& 54.2& 60.9& 59.5& 66.2& 74.6\\
(15\_50\_02\_L)& (86.9)& (88.8)& (86.0)& (85.1)& (71.8)& (74.5)& (53.8)& (62.3)& (66.4)\\
15\_100\_STR& 0.0& 0.3& 0.5& 0.0& 0.0& 0.0& 1.0& 2.7& 2.3\\
15\_100\_1& 85.5& 81.3& 29.2& 68.8& 74.0& 15.0& 81.0& 86.0& 8.7\\
15\_100\_02& 92.2& 91.7& 75.3& 83.5& 74.2& 70.7& 63.3& 56.0& 75.7\\
30\_50\_STR& 0.2& 0.8& 1.0& 0.0& 0.0& 0.0& 1.3& 3.0& 2.7\\
30\_50\_1& 93.2& 94.0& 79.2& 73.1& 69.1& 40.3& 88.0& 93.0& 63.2\\
30\_50\_02& 91.5& 96.8& 91.0& 71.9& 36.3& 37.0& 66.2& 44.5& 49.5\\
(30\_50\_02\_L)& (89.5)& (95.7)& (90.1)& (84.4)& (61.0)& (60.8)& (64.1)& (44.6)& (47.7)\\
30\_100\_STR& 0.3& 0.2& 0.5& 0.0& 0.0& 0.0& 0.7& 1.7& 1.3\\
30\_100\_1& 94.0& 92.5& 71.0& 75.0& 75.5& 55.3& 90.0& 91.7& 38.3\\
30\_100\_02& 92.5& 91.5& 90.8& 91.3& 74.7& 71.7& 74.7& 56.0& 64.0\\
(30\_100\_02\_L)& (96.2)& (94.4)& (92.7)& (95.7)& (85.4)& (85.8)& (75.2)& (63.2)& (65.2)\\
45\_50\_STR& 0.5& 1.5& 1.5& 0.0& 0.0& 0.0& 0.7& 1.0& 2.3\\
45\_50\_1& 96.2& 95.2& 71.7& 78.8& 72.2& 38.1& 92.3& 94.3& 55.5\\
45\_50\_02& 94.0& 95.0& 77.3& 49.3& 36.1& 48.7& 50.5& 37.8& 51.2\\
(45\_50\_02\_L)& (93.5)& (94.7)& (73.4)& (73.9)& (63.1)& (70.1)& (54.6)& (41.6)& (48.1)\\
45\_100\_STR& 0.5& 0.8& 0.3& 0.0& 0.0& 0.0& 0.3& 0.7& 0.3\\
45\_100\_1& 95.7& 95.0& 86.5& 82.8& 78.5& 48.3& 93.0& 95.3& 43.0\\
45\_100\_02& 95.3& 95.5& 77.7& 77.3& 86.5& 69.7& 51.3& 77.0& 66.0\\
(45\_100\_02\_L)& (96.5)& (96.6)& (81.5)& (88.7)& (93.2)& (84.8)& (61.5)& (73.5)& (62.0)\\

\hline
\end{tabular}
\end{table*}

\subsection{Resolution dependence}
\label{subsec:resolution}
\begin{figure*}
    \centering
    \includegraphics[width=\linewidth]{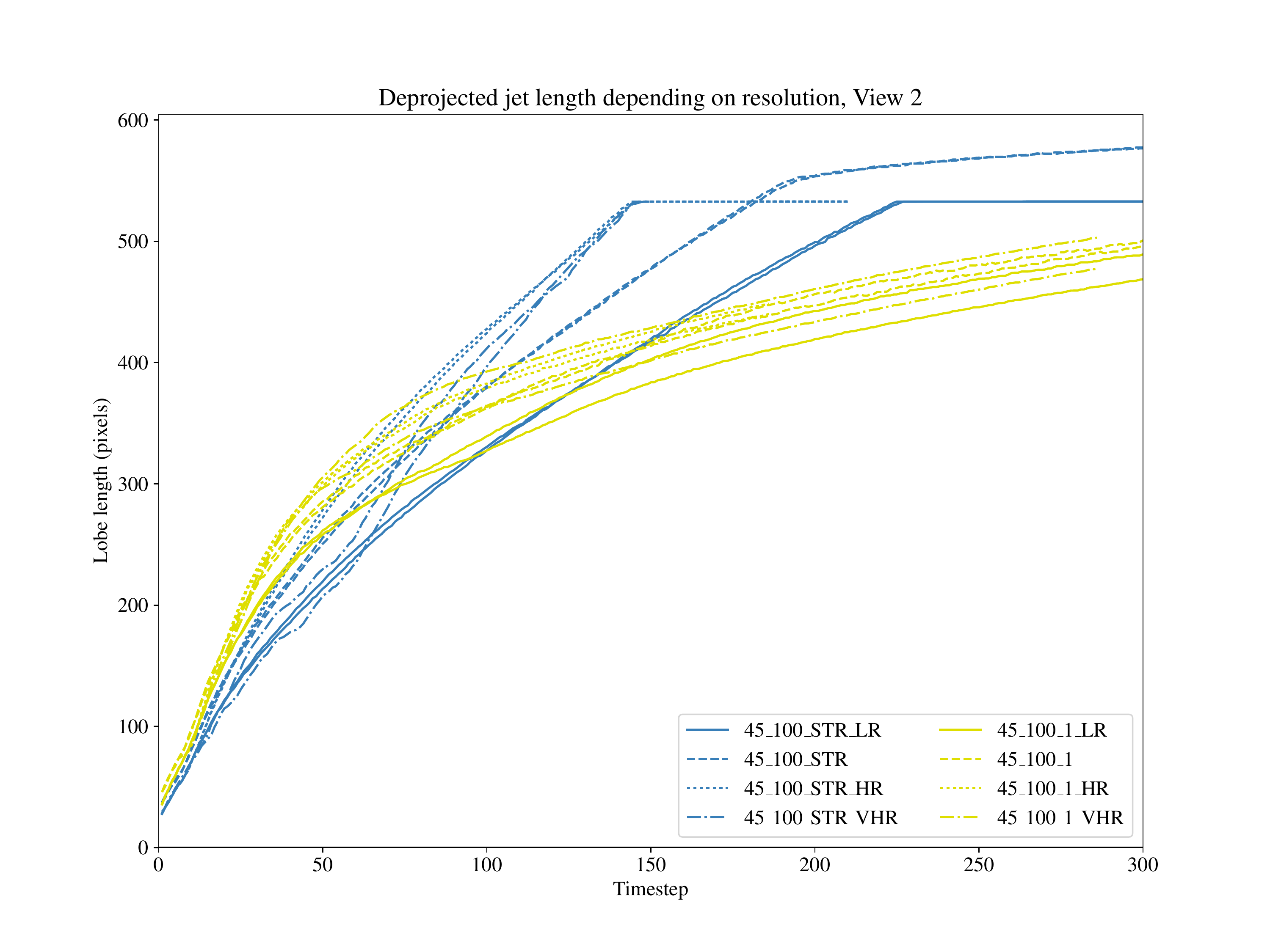}
    \caption{Projection-corrected lobe length resolution dependency plot for `View 2' of precessing and straight jets. Blue corresponds to non-precessing jets whilst yellow signifies precessing ones; straight lines are for low resolution (LR) jets, dashed lines are for the original runs, dotted lines are high resolution (HR) and dashed and dotted lines are very high resolution (VHR).}
    \label{fig:res_jetgrowth}
\end{figure*}

\begin{figure*}
    \centering
    \includegraphics[width=\linewidth]{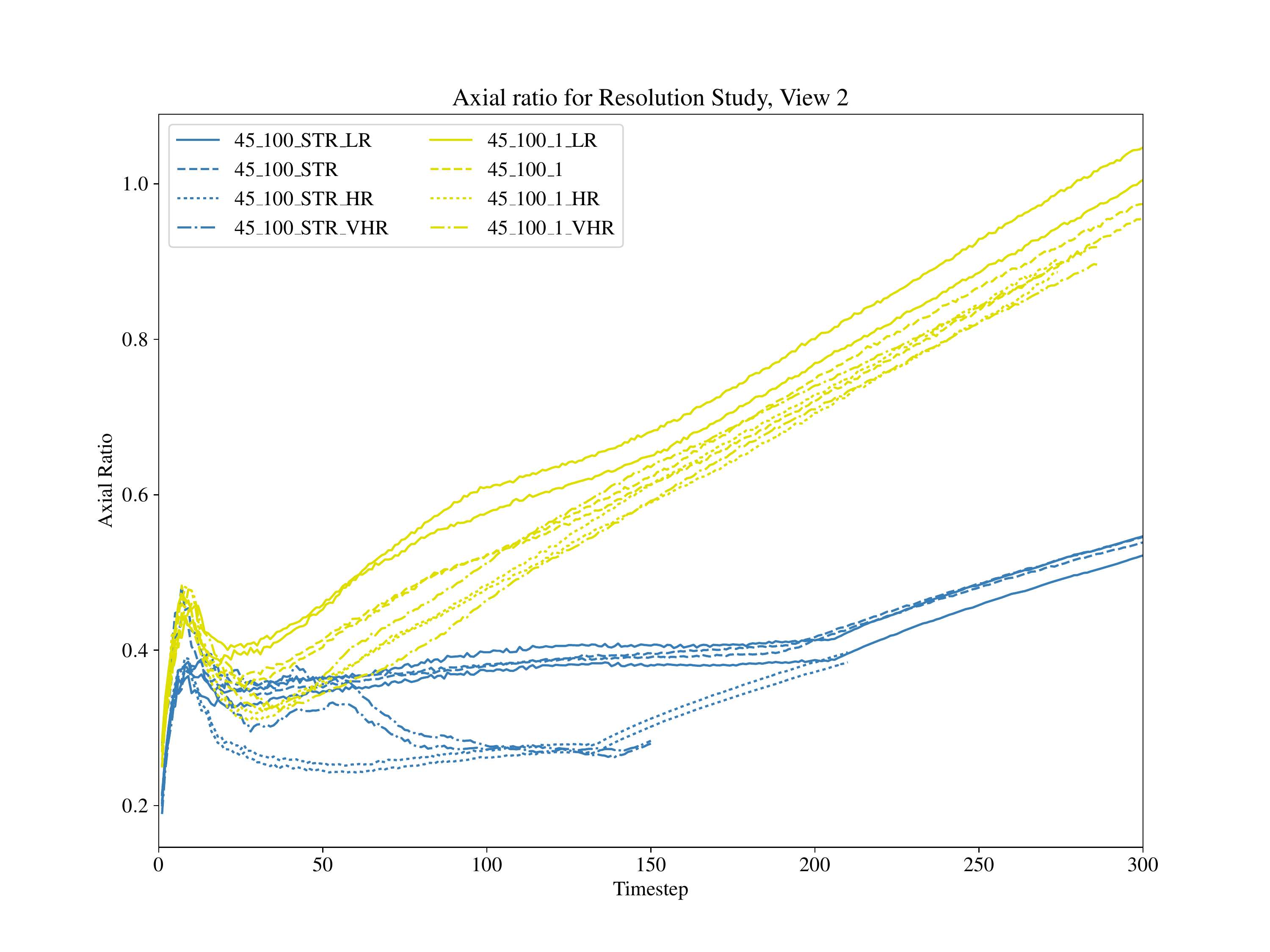}
    \caption{Axial ratio changes for `View 2' of precessing straight jets. Colours and linestyles as in Fig.~\ref{fig:res_jetgrowth}.}
    \label{fig:res_aspect}
\end{figure*}

\begin{figure*}
    \centering
    \includegraphics[width=\linewidth]{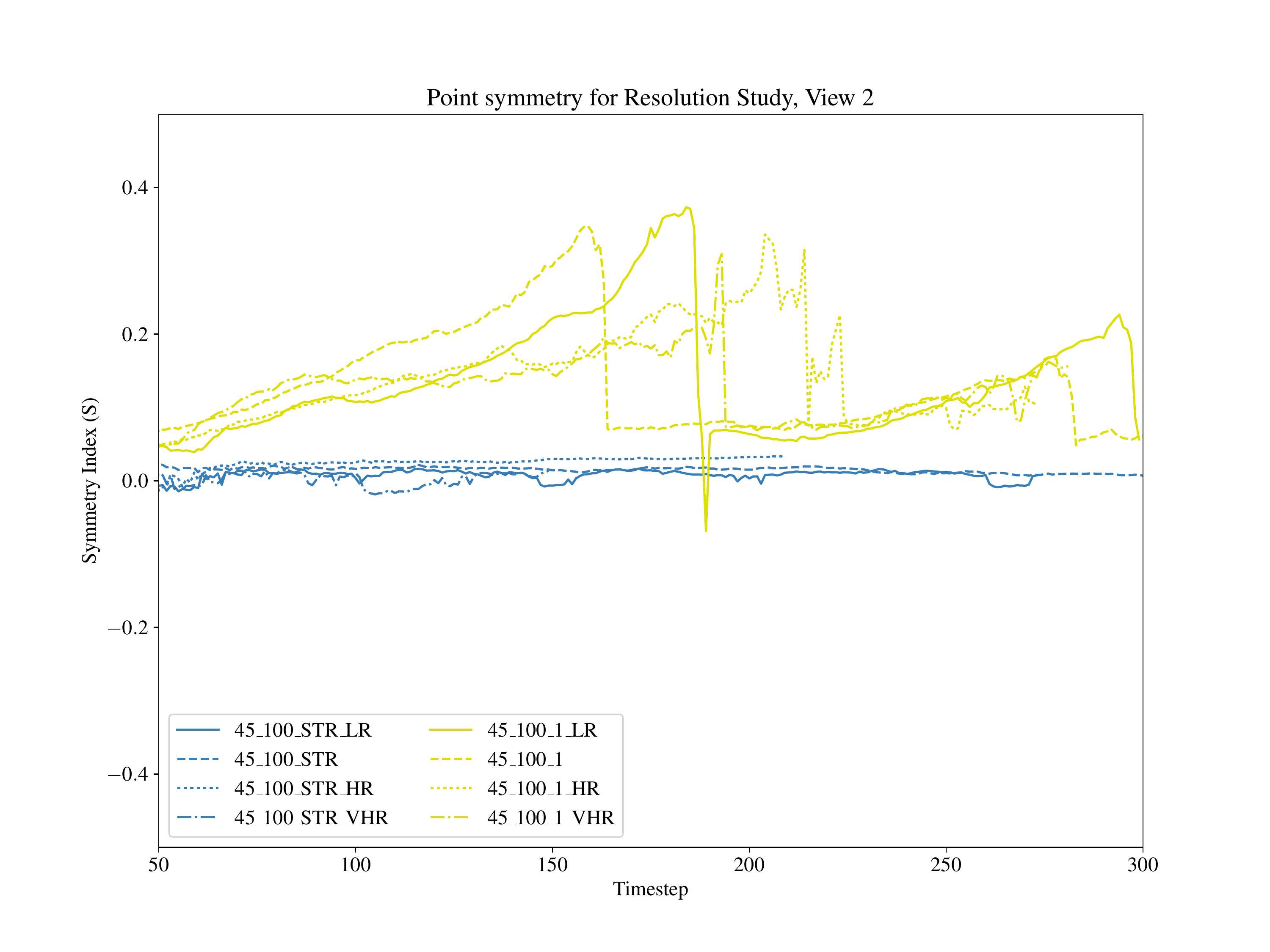}
    \caption{Resolution dependence on point symmetry precession indicator (S), for `View 2' of simulation runs. Colours and linestyles as in Fig.~\ref{fig:res_jetgrowth}.}
    \label{fig:res_sness.pdf}
\end{figure*}

\begin{figure*}
    \centering
    \includegraphics[width=\linewidth]{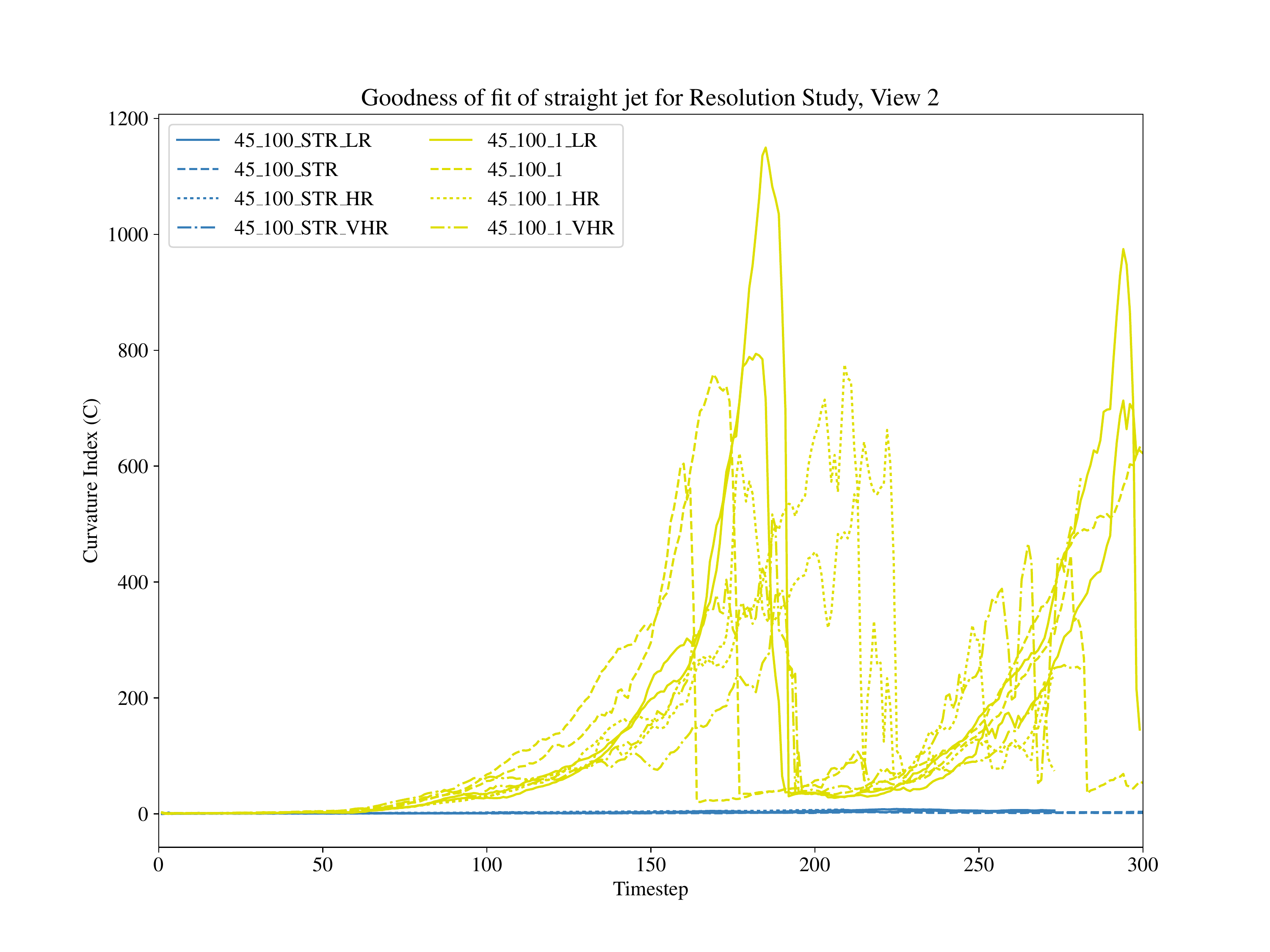}
    \caption{Resolution dependence on curvature precession indicator (C), for `View 2' of simulation runs. Colours and linestyles as in Fig.~\ref{fig:res_jetgrowth}.}
    \label{fig:res_curvature}
\end{figure*}

\begin{figure*}
    \centering
    \includegraphics[width=\linewidth]{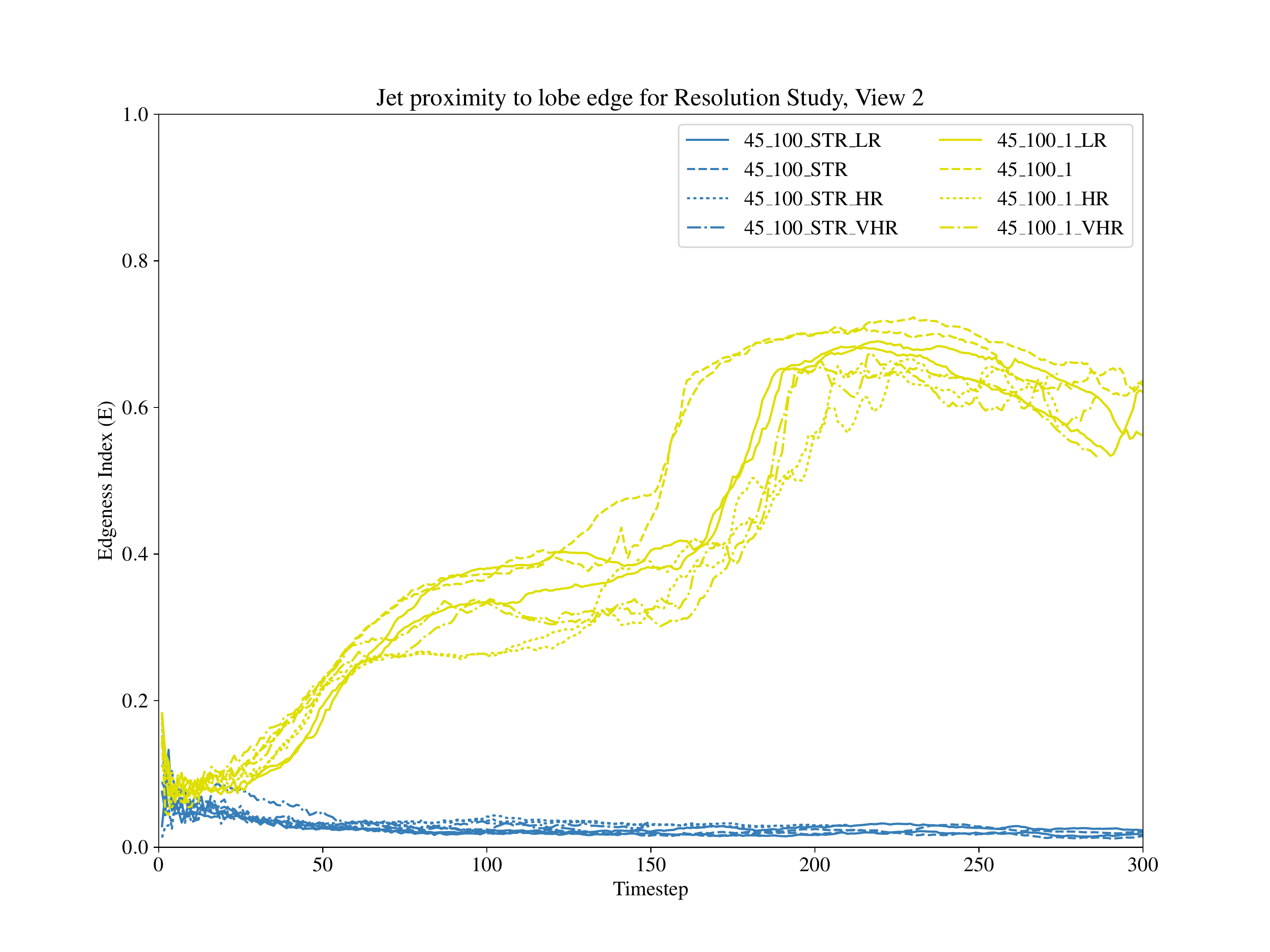}
    \caption{Resolution dependence on edgeness precession indicator (E), for `View 2' of simulation runs. Colours and linestyles as in Fig.~\ref{fig:res_jetgrowth}.}
    \label{fig:res_edgeness}
\end{figure*}

Figures \ref{fig:res_jetgrowth} to \ref{fig:res_edgeness} show the influence of resolution on both straight and precessing jets. The precessing jet was chosen for having a wide cone angle, slower precession period and a high Mach number (see Fig.~\ref{fig:gridview}, row 2, column 2), allowing for dynamic morphologies without the complex interactions seen at higher precession periods. 

For the lobe environment morphology indicators (lobe growth and axial ratio, Figures \ref{fig:res_jetgrowth} and \ref{fig:res_aspect}), the precessing jets (yellow) show some spread in lobe growth and width but the overall behaviours are consistent throughout the simulation. 

\begin{table*} 
\caption{Fraction of simulation time where jet shows precession indicators, resolution study.}
\label{table:curve_res}
\centering
\begin{tabular}{lrrrrrrrrr}
\hline
Simulation name & \multicolumn{3}{c}{Jet at edge (E)} & \multicolumn{3}{c}{Jet curved (C)} &\multicolumn{3}{c}{Jet S-symmetric (S)}\\
& View 1 & View 2 & View 3 & View 1 & View 2 & View 3 & View 1 & View 2 & View 3   \\
\hline
45\_100\_STR\_LR& 0.0& 0.3& 0.0& 0.0& 0.0& 0.0& 0.7& 2.2& 1.1\\
45\_100\_STR& 0.5& 0.8& 0.3& 0.0& 0.0& 0.0& 0.3& 0.7& 0.3\\
45\_100\_STR\_HR& 0.0& 0.2& 0.0& 0.0& 0.0& 0.0& 1.0& 3.8& 1.4\\
45\_100\_STR\_VHR& 0.0& 0.7& 0.0& 0.3& 0.0& 0.0& 1.3& 6.0& 1.3\\
45\_100\_1\_LR& 95.5& 90.8& 86.2& 78.4& 76.1& 49.7& 90.6& 85.6& 46.8\\
45\_100\_1& 95.7& 95.0& 86.5& 82.8& 78.5& 48.3& 93.0& 95.3& 43.0\\
45\_100\_1\_HR& 93.8& 91.4& 73.7& 77.1& 73.8& 69.6& 94.1& 91.9& 54.2\\
45\_100\_1\_VHR& 94.8& 93.5& 83.6& 80.8& 76.9& 72.8& 94.0& 91.8& 73.3\\
\hline
\end{tabular}
\end{table*}

\section{Discussion}
\subsection{Resolution dependence and model validation}
The resolution study described in Section~\ref{subsec:resolution} examines the impact of resolution on both lobe structure and precession markers. Figs.~\ref{fig:res_jetgrowth} and \ref{fig:res_aspect} show the changes throughout each simulation. Surprisingly, it is the straight jets which are less uniform, with lower resolution jets being more stable -- e.g., less likely to break up -- over time. Not only do higher resolution straight jets run off the grid faster (e.g., 45\_100\_STR\_VHR has only 150 on-grid timesteps), but axial ratio decreases as a consequence of faster growth. The dependence of source expansion speed on resolution is opposite from what is expected from resolution studies of pre-collimated jets \citep{krause01}. This is probably due to a resolution dependence in the hydrodynamic collimation process that we include self-consistently here by injecting jets with a finite opening angle.

When the precession indicators themselves are examined, these roles reverse. Figures~ \ref{fig:res_sness.pdf} (S), \ref{fig:res_curvature} (C), and \ref{fig:res_edgeness} (E) show that the straight jet continues to be straight at all resolutions with the edgeness (E) indicator showing a slight increase in straight jets from expansion, and point symmetry (S) showing some spread in higher resolution jets which are typically less stable. Yet these account for only minor variations in indicator scoring, as shown in Table~\ref{table:curve_res}. 

It is interesting to note that resolution dependency is \textit{also} dependent upon simulation view. For example, edgeness (E) indicators are very similar for the precessing jet for Views 1 and 2, which are the more curved projections. Yet for View 3, which typically has fewer indicators, the effects of resolution over time are slightly reduced. On one hand, lower resolution jets witness a reduction in the internal hydrodynamics which drive precession markers; on the other, higher resolution jets are less stable and break up more often, leading to worse fits in some projections. However, both high and low resolution jets pass indicator thresholds for roughly the same amount of time for all three indicators in all views.

We conclude that the prevalence of precession indicators beyond each threshold is essentially independent of resolution and so it is safe to analyse these indicators based on our parameter study. 

\subsection{Precession indicators and jet structure}

It is immediately apparent that generally all simulations of precessing jets exhibit one or more of the precession markers indicated by \cite{krause18} at most times. Table~\ref{table:multi_indicator} shows that some combinations of parameters, specifically those with narrow precession cone angles and longer precession periods, resemble straight jets for much of their runs. Yet even for the simulations and views which show the least visible precession, a marker can be detected approximately one third of the time (Table~\ref{table:multi_indicator}), and usually more, whilst no false positive is present for more than 5\% of the time. 

Of course, the sources where precession is harder to detect have consequences observationally; in observations a single source is only ever viewed as a snapshot, and as such will have a random orientation which may show little precession. This can be caused by an ambiguity during the first precession turn where it depends strongly on the location of the observer relative to the source, if it will appear to be precessing or not. From certain views the source will not have precessed enough relative to the observer for indicators to appear, or the curvature in the jet may be hidden by projection. But if even one precession indicator is observed in a source, it is almost certainly caused by real precession rather than hydrodynamics alone. 

When two markers are present, the rate of false positives is 0\% for most straight jets; this increases only very slightly over certain views and is accounted for by early expansion. Clearly, it is the rate of false negatives -- where orientation and internal dynamics align to cause real signatures to be undercounted -- which is far more of a problem. Many jets appear straight whilst actually precessing: this can be seen easily in Fig.~\ref{fig:gridview}, particularly in column 3, and again in Fig.~\ref{fig:gridview-45_100_02_L} where (S) values are often low despite the complex lobe morphologies which are naturally formed from high precession periods and wide precession cone opening angles. In these cases, despite being rapidly precessing, and having complex lobe hydrodynamics as a consequence, the jets themselves often appear visually straight and show low values for curvature and point symmetry. The fact that certain populations of precessing jets (especially the long-period jets) may not be detected as precessing from certain views at certain times means that the binary population of potentially precessing real-world sources given by \cite{krause18} may well be underestimated.

Averaging over simulations with different cone opening angles and Mach numbers for our simulations with well developed precession (5 turns simulated, \_02 runs), we find that the probability that a precessing source shows at least one precession indicators is 77\%, whilst two and three indicators are both 72\%. Looking at the same source parameters for the straight jets (\_STR runs), we find the average presence of one, two and three precession indicators are 1.2\%, 1.7\% and 1.7\% of the time respectively. Defining a sample of all analysed snapshots for all simulations with well developed precession and also with no precession (i.e., the \_02 runs and \_STR runs together), we find that a randomly drawn simulated radio source with precession indicators has a 98\% chance of being a precessing source, and this is the same no matter how many precession indicators are present. In other words, the morphological precession indicators, where present, are very reliable indicators of true precession. This result is obviously within the framework of our simulations. We have not taken into account triaxial dark matter halos \citep{rossi17}, or sloshing of the intra-group / cluster medium \citep[e.g.,][]{werner10}, which could plausibly enhance precession indicators for straight jets.

\begin{table*} 
\caption{Percentage of time where simulations show at least one indicator (I = 1), at least two indicators (I = 2) or all three (I = 3), for each of the three views.}
\label{table:multi_indicator}
\centering
\begin{tabular}{lrrrrrrrrr}
\hline
Simulation name & \multicolumn{3}{c}{I = 1} & \multicolumn{3}{c}{I = 2} &\multicolumn{3}{c}{I = 3}\\
& View 1 & View 2 & View 3 & View 1 & View 2 & View 3 & View 1 & View 2 & View 3   \\
\hline
15\_50\_STR& 3.0 & 0.0 & 0.0 & 4.3 & 1.7 & 0.0 & 5.0 & 1.0 & 0.0 \\
15\_50\_1& 85.7 & 59.0 & 31.0 & 85.0 & 64.0 & 27.7 & 35.3 & 12.0 & 7.3 \\
15\_50\_02& 96.3 & 82.7 & 50.0 & 98.0 & 74.7 & 51.0 & 94.0 & 71.0 & 52.7 \\
(15\_50\_02\_L)& (98.2) & (90.5) & (47.0) & (99.0) & (85.8) & (52.2) & (97.0) & (81.3) & (55.0) \\
15\_100\_STR& 2.0 & 0.0 & 0.0 & 3.3 & 0.7 & 0.0 & 2.7 & 1.0 & 0.0 \\
15\_100\_1& 82.3 & 71.3 & 30.0 & 86.0 & 77.0 & 24.7 & 39.0 & 13.0 & 0.0 \\
15\_100\_02& 95.7 & 86.7 & 51.3 & 97.7 & 83.7 & 39.3 & 92.0 & 80.7 & 52.3 \\
30\_50\_STR& 4.0 & 1.0 & 0.0 & 6.0 & 0.7 & 0.0 & 6.0 & 0.7 & 0.0 \\
30\_50\_1& 91.7 & 74.3 & 64.0 & 96.0 & 73.0 & 62.3 & 83.3 & 56.0 & 32.7 \\
30\_50\_02& 99.7 & 87.0 & 56.7 & 99.7 & 56.3 & 31.3 & 99.0 & 58.7 & 33.0 \\
(30\_50\_02\_L)& (99.8) & (91.8) & (56.3) & (99.8) & (74.3) & (36.2) & (99.5) & (77.0) & (38.0) \\
30\_100\_STR& 2.0 & 0.3 & 0.0 & 2.0 & 1.0 & 0.0 & 2.0 & 1.0 & 0.0 \\
30\_100\_1& 92.3 & 78.3 & 64.3 & 95.3 & 78.7 & 59.7 & 75.7 & 63.3 & 25.3 \\
30\_100\_02& 98.3 & 94.3 & 69.3 & 99.0 & 85.0 & 48.7 & 96.3 & 80.3 & 56.0 \\
(30\_100\_02\_L)& (99.2) & (95.7) & (69.0) & (99.5) & (91.8) & (58.2) & (98.2) & (90.2) & (59.5) \\
45\_50\_STR& 6.3 & 0.3 & 0.0 & 7.0 & 0.3 & 0.0 & 7.0 & 2.0 & 0.0 \\
45\_50\_1& 98.0 & 84.3 & 69.7 & 97.0 & 80.7 & 68.7 & 91.0 & 55.3 & 29.3 \\
45\_50\_02& 99.3 & 64.7 & 37.0 & 99.7 & 53.0 & 23.3 & 99.3 & 66.3 & 32.7 \\
(45\_50\_02\_L)& (99.7) & (80.7) & (44.2) & (99.8) & (74.2) & (33.3) & (99.7) & (81.0) & (37.7) \\
45\_100\_STR& 2.7 & 0.3 & 0.0 & 2.7 & 0.7 & 0.0 & 2.7 & 0.3 & 0.0 \\
45\_100\_1& 94.3 & 83.7 & 76.0 & 96.7 & 81.7 & 72.3 & 83.7 & 55.3 & 28.3 \\
45\_100\_02& 99.3 & 84.7 & 45.0 & 98.3 & 91.7 & 69.0 & 97.3 & 83.0 & 49.0 \\
(45\_100\_02\_L)& (99.7) & (89.7) & (56.8) & (99.2) & (94.7) & (68.8) & (98.7) & (90.5) & (52.7) \\

\hline
\end{tabular}
\end{table*}

\subsection{Precession effects and lobe dynamics}

Figures \ref{fig:lobelength_allsims} and \ref{fig:lobegrowth_allsims} show that precession has a strong effect on lobe dynamics.  The wider the precession cone opening angle, the slower the source expansion, and the more rapidly the jet is precessing, the less growth there is. 

Interestingly, the influence of precession cone opening angle is comparatively minor: between our $15^\circ$ and our $45^\circ$ simulations, the expansion rate changes only by about 20\%. In contrast, even the slowly precessing jets expand about $1/3$ slower than the non-precessing ones, with our fast-precessing sources at roughly 50 per cent of the straight-source speed. This could have important consequences for jet power determinations (compare, e.g., \citealt{turner18,hardcastle19b}), as jet power is strongly dependent on source size in these models. 

\subsection{Morphological comparisons to real-world sources}
Many of the simulation snapshots are comparable to complex morphologies found in current observations of X-ray and radio jets. For slow precession periods, precessing jets can give rise to lobe morphologies comparable to those of X- or Z-shaped sources (Fig. \ref{fig:gridview}), which have in the past been attributed to rapid jet reorientation as a result of black hole-black hole mergers, \citep{merritt02} to complex structures in the host environment \citep{leahy84,hardcastle19}, or to hydrodynamic backflow \citep{cotton20}. Importantly, these structures can exist while the jets appear straight, so jet precession as an origin of X-shaped sources cannot be ruled out on the basis of jets appearing straight. 

Jets where the precession period is short compared to the source lifetime make complex amorphous morphologies that are perhaps less well matched to typical sources, particularly when the precession angle is also large, although some of the structures produced are reminiscent of restarting or `double-double' sources \citep{schoenmakers00}, or hybrid sources \citep{harwood20}. It may be that jets with short precession periods are rare or short-lived, but further work would be required to explore this. The morphologies shown in the late stages of Fig.~\ref{fig:gridview-45_100_02_L} may be unlikely to be observed in reality given likely disturbances from the intra-cluster / intra-group medium over the dynamical age of the host galaxy. 

\section{Conclusions and future work}
We have simulated radio sources with precessing jets and produced synthetic radio images with separate proxies for lobe and jet emission. We have found that:

\begin{itemize}
    \item Physical properties of the precessing jet system are responsible for complex morphologies which mimic structures observed in real-world radio sources.
    \item Jet precession results in predictable changes to jet and lobe structures, which become more pronounced with extreme precession cone opening angles and faster precession periods. 
    \item Fast and slow precessing jets result in changes to jet curvature and stability, with rapidly precessing jets breaking up more often while slowly precessing jets can show smooth curvature.
    \item Jet velocity has little impact on morphology beyond slower jets producing more compact sources.
    \item Precessing jets often appear straight, and certain viewing angles and physical properties make it very difficult to detect precession markers. 
    \item Source expansion slows down significantly with decreasing precession period. Our fast-precessing jets expand at merely 50\% of the speed of the equivalent straight-jet source. This will have a strong impact on the determination of the jet power of such a radio source. 
    \item All three investigated precession markers are useful for classifying precessing jets. Each one is present in precessing sources most of the time and in non-precessing sources almost never. Somewhat exceptional is S-symmetry, which is present in straight-jet sources in a few per cent of the investigated snapshots.
    \item If a source displays one or more precession markers according to our definitions, the overall probability is 98\% that the source hosts a precessing jet. Hence, any radio source that shows either S-symmetry or a misaligned jet at the edge of the lobe or significantly curved jets is very likely precessing. 
    \item Since real-world observations necessarily involve a single snapshot at a single point in a source's lifetime, this may lead to an underestimation of the number of active supermassive black hole binaries producing precessing jets.
\end{itemize}

Future work will look at jet hydrodynamical processes and MCMC jet path fitting to observed and simulated sources.




\section*{Acknowledgements}

MAH acknowledges a studentship from STFC [ST/R504786/1] and MJH acknowledges support from STFC [ST/R000905/1]. This research made use of the University of Hertfordshire high-performance computing facility (\url{https://uhhpc.herts.ac.uk/}).




\section*{Data availability}
No new observational data were generated or analysed in support of this research. Simulation source files are available on request. 

\bibliographystyle{mnras}
\bibliography{main} 

\begin{thebibliography}{}
\makeatletter
\relax
\def\mn@urlcharsother{\let\do\@makeother \do\$\do\&\do\#\do\^\do\_\do\%\do\~}
\def\mn@doi{\begingroup\mn@urlcharsother \@ifnextchar [ {\mn@doi@}
  {\mn@doi@[]}}
\def\mn@doi@[#1]#2{\def\@tempa{#1}\ifx\@tempa\@empty \href
  {http://dx.doi.org/#2} {doi:#2}\else \href {http://dx.doi.org/#2} {#1}\fi
  \endgroup}
\def\mn@eprint#1#2{\mn@eprint@#1:#2::\@nil}
\def\mn@eprint@arXiv#1{\href {http://arxiv.org/abs/#1} {{\tt arXiv:#1}}}
\def\mn@eprint@dblp#1{\href {http://dblp.uni-trier.de/rec/bibtex/#1.xml}
  {dblp:#1}}
\def\mn@eprint@#1:#2:#3:#4\@nil{\def\@tempa {#1}\def\@tempb {#2}\def\@tempc
  {#3}\ifx \@tempc \@empty \let \@tempc \@tempb \let \@tempb \@tempa \fi \ifx
  \@tempb \@empty \def\@tempb {arXiv}\fi \@ifundefined
  {mn@eprint@\@tempb}{\@tempb:\@tempc}{\expandafter \expandafter \csname
  mn@eprint@\@tempb\endcsname \expandafter{\@tempc}}}

\bibitem[\protect\citeauthoryear{{Begelman}, {Blandford}  \& {Rees}}{{Begelman}
  et~al.}{1980}]{begelman80}
{Begelman} M.~C.,  {Blandford} R.~D.,   {Rees} M.~J.,  1980, \mn@doi [\nat]
  {10.1038/287307a0}, \href {http://adsabs.harvard.edu/abs/1980Natur.287..307B}
  {287, 307}

\bibitem[\protect\citeauthoryear{{Chandrasekhar}}{{Chandrasekhar}}{1961}]{chandra03}
{Chandrasekhar} S.,  1961, {Hydrodynamic and hydromagnetic stability}.
Oxford University Press

\bibitem[\protect\citeauthoryear{{Cotton} et~al.,}{{Cotton}
  et~al.}{2020}]{cotton20}
{Cotton} W.~D.,  et~al., 2020, \mn@doi [\mnras] {10.1093/mnras/staa1240}, \href
  {https://ui.adsabs.harvard.edu/abs/2020MNRAS.tmp.1389C} {}

\bibitem[\protect\citeauthoryear{{Cox}, {Gull}  \& {Scheuer}}{{Cox}
  et~al.}{1991}]{cox91}
{Cox} C.~I.,  {Gull} S.~F.,   {Scheuer} P.~A.~G.,  1991, \mn@doi [\mnras]
  {10.1093/mnras/252.4.558}, \href
  {https://ui.adsabs.harvard.edu/abs/1991MNRAS.252..558C} {252, 558}

\bibitem[\protect\citeauthoryear{{Donohoe} \& {Smith}}{{Donohoe} \&
  {Smith}}{2016}]{donohoe16}
{Donohoe} J.,  {Smith} M.~D.,  2016, \mn@doi [\mnras] {10.1093/mnras/stw335},
  \href {https://ui.adsabs.harvard.edu/abs/2016MNRAS.458..558D} {458, 558}

\bibitem[\protect\citeauthoryear{{English}, {Hardcastle}  \&
  {Krause}}{{English} et~al.}{2016}]{english16}
{English} W.,  {Hardcastle} M.~J.,   {Krause} M.~G.~H.,  2016, \mn@doi [\mnras]
  {10.1093/mnras/stw1407}, \href
  {https://ui.adsabs.harvard.edu/abs/2016MNRAS.461.2025E} {461, 2025}

\bibitem[\protect\citeauthoryear{{Gower}, {Gregory}, {Unruh}  \&
  {Hutchings}}{{Gower} et~al.}{1982}]{gower82}
{Gower} A.~C.,  {Gregory} P.~C.,  {Unruh} W.~G.,   {Hutchings} J.~B.,  1982,
  \mn@doi [\apj] {10.1086/160442}, \href
  {http://cdsads.u-strasbg.fr/abs/1982ApJ...262..478G} {262, 478}

\bibitem[\protect\citeauthoryear{{Hardcastle} \& {Krause}}{{Hardcastle} \&
  {Krause}}{2013}]{hardcastle13}
{Hardcastle} M.~J.,  {Krause} M.~G.~H.,  2013, \mn@doi [\mnras]
  {10.1093/mnras/sts564}, \href
  {http://adsabs.harvard.edu/abs/2013MNRAS.430..174H} {430, 174}

\bibitem[\protect\citeauthoryear{{Hardcastle} et~al.,}{{Hardcastle}
  et~al.}{2016}]{hardcastle16}
{Hardcastle} M.~J.,  et~al., 2016, \mn@doi [\mnras] {10.1093/mnras/stv2553},
  \href {https://ui.adsabs.harvard.edu/abs/2016MNRAS.455.3526H} {455, 3526}

\bibitem[\protect\citeauthoryear{{Hardcastle} et~al.,}{{Hardcastle}
  et~al.}{2019a}]{hardcastle19}
{Hardcastle} M.~J.,  et~al., 2019a, \mn@doi [Monthly Notices of the Royal
  Astronomical Society] {10.1093/mnras/stz1910}, \href
  {https://ui.adsabs.harvard.edu/abs/2019MNRAS.488.3416H} {488, 3416}

\bibitem[\protect\citeauthoryear{{Hardcastle} et~al.,}{{Hardcastle}
  et~al.}{2019b}]{hardcastle19b}
{Hardcastle} M.~J.,  et~al., 2019b, \mn@doi [\aap]
  {10.1051/0004-6361/201833893}, \href
  {https://ui.adsabs.harvard.edu/abs/2019A&A...622A..12H} {622, A12}

\bibitem[\protect\citeauthoryear{{Harwood}, {Vernstrom}  \& {Stroe}}{{Harwood}
  et~al.}{2020}]{harwood20}
{Harwood} J.~J.,  {Vernstrom} T.,   {Stroe} A.,  2020, \mn@doi [\mnras]
  {10.1093/mnras/stz3069}, \href
  {https://ui.adsabs.harvard.edu/abs/2020MNRAS.491..803H} {491, 803}

\bibitem[\protect\citeauthoryear{{Horton}, {Hardcastle}, {Read}  \&
  {Krause}}{{Horton} et~al.}{2020}]{horton20a}
{Horton} M.~A.,  {Hardcastle} M.~J.,  {Read} S.~C.,   {Krause} M. G.~H.,  2020,
  \mn@doi [\mnras] {10.1093/mnras/staa429}, \href
  {https://ui.adsabs.harvard.edu/abs/2020MNRAS.tmp..407H} {}

\bibitem[\protect\citeauthoryear{{Krause} \& {Camenzind}}{{Krause} \&
  {Camenzind}}{2001}]{krause01}
{Krause} M.,  {Camenzind} M.,  2001, \mn@doi [\aap]
  {10.1051/0004-6361:20011452}, \href
  {https://ui.adsabs.harvard.edu/abs/2001A&A...380..789K} {380, 789}

\bibitem[\protect\citeauthoryear{{Krause}, {Alexander}, {Riley}  \&
  {Hopton}}{{Krause} et~al.}{2012}]{krause12}
{Krause} M.,  {Alexander} P.,  {Riley} J.,   {Hopton} D.,  2012, \mn@doi
  [\mnras] {10.1111/j.1365-2966.2012.21645.x}, \href
  {https://ui.adsabs.harvard.edu/abs/2012MNRAS.427.3196K} {427, 3196}

\bibitem[\protect\citeauthoryear{{Krause} et~al.,}{{Krause}
  et~al.}{2019}]{krause18}
{Krause} M. G.~H.,  et~al., 2019, \mn@doi [\mnras] {10.1093/mnras/sty2558},
  \href {https://ui.adsabs.harvard.edu/abs/2019MNRAS.482..240K} {482, 240}

\bibitem[\protect\citeauthoryear{{Leahy} \& {Williams}}{{Leahy} \&
  {Williams}}{1984}]{leahy84}
{Leahy} J.~P.,  {Williams} A.~G.,  1984, \mn@doi [\mnras]
  {10.1093/mnras/210.4.929}, \href
  {https://ui.adsabs.harvard.edu/abs/1984MNRAS.210..929L} {210, 929}

\bibitem[\protect\citeauthoryear{{Mayer}}{{Mayer}}{2017}]{mayer17}
{Mayer} L.,  2017, in Journal of Physics Conference Series. p. 012025
  (\mn@eprint {arXiv} {1703.00661}), \mn@doi{10.1088/1742-6596/840/1/012025}

\bibitem[\protect\citeauthoryear{{Merritt} \& {Ekers}}{{Merritt} \&
  {Ekers}}{2002}]{merritt02}
{Merritt} D.,  {Ekers} R.~D.,  2002, \mn@doi [Science]
  {10.1126/science.1074688}, \href
  {https://ui.adsabs.harvard.edu/abs/2002Sci...297.1310M} {297, 1310}

\bibitem[\protect\citeauthoryear{{Mignone}, {Bodo}, {Massaglia}, {Matsakos},
  {Tesileanu}, {Zanni}  \& {Ferrari}}{{Mignone} et~al.}{2007}]{mignone07}
{Mignone} A.,  {Bodo} G.,  {Massaglia} S.,  {Matsakos} T.,  {Tesileanu} O.,
  {Zanni} C.,   {Ferrari} A.,  2007, in JENAM-2007, ``Our Non-Stable
  Universe''. pp 96--96

\bibitem[\protect\citeauthoryear{{Rossi}, {Marchetti}, {Cacciato}, {Kuiack}  \&
  {Sari}}{{Rossi} et~al.}{2017}]{rossi17}
{Rossi} E.~M.,  {Marchetti} T.,  {Cacciato} M.,  {Kuiack} M.,   {Sari} R.,
  2017, \mn@doi [\mnras] {10.1093/mnras/stx098}, \href
  {https://ui.adsabs.harvard.edu/abs/2017MNRAS.467.1844R} {467, 1844}

\bibitem[\protect\citeauthoryear{{Schoenmakers}, {de Bruyn}, {R{\"o}ttgering},
  {van der Laan}, {Mack}  \& {Kaiser}}{{Schoenmakers}
  et~al.}{2000}]{schoenmakers00}
{Schoenmakers} A.~P.,  {de Bruyn} A.~G.,  {R{\"o}ttgering} H.~J.~A.,  {van der
  Laan} H.,  {Mack} K.~H.,   {Kaiser} C.~R.,  2000, in {van Haarlem} M.~P.,
  ed., Perspectives on Radio Astronomy: Science with Large Antenna Arrays.
  p.~165 (\mn@eprint {arXiv} {astro-ph/9910448})

\bibitem[\protect\citeauthoryear{{Smith} \& {Donohoe}}{{Smith} \&
  {Donohoe}}{2019}]{smith19}
{Smith} M.~D.,  {Donohoe} J.,  2019, \mn@doi [\mnras] {10.1093/mnras/stz2525},
  \href {https://ui.adsabs.harvard.edu/abs/2019MNRAS.490.1363S} {490, 1363}

\bibitem[\protect\citeauthoryear{{Sun}, {Yang}, {Rieger}, {Liu}  \&
  {Aharonian}}{{Sun} et~al.}{2018}]{sun18}
{Sun} X.-N.,  {Yang} R.-Z.,  {Rieger} F.~M.,  {Liu} R.-Y.,   {Aharonian} F.,
  2018, \mn@doi [\aap] {10.1051/0004-6361/201731716}, \href
  {https://ui.adsabs.harvard.edu/abs/2018A&A...612A.106S} {612, A106}

\bibitem[\protect\citeauthoryear{{Tremmel}, {Governato}, {Volonteri}, {Pontzen}
   \& {Quinn}}{{Tremmel} et~al.}{2018}]{tremmel18}
{Tremmel} M.,  {Governato} F.,  {Volonteri} M.,  {Pontzen} A.,   {Quinn} T.~R.,
   2018, \mn@doi [\apjl] {10.3847/2041-8213/aabc0a}, \href
  {http://cdsads.u-strasbg.fr/abs/2018ApJ...857L..22T} {857, L22}

\bibitem[\protect\citeauthoryear{{Turner}, {Shabala}  \& {Krause}}{{Turner}
  et~al.}{2018}]{turner18}
{Turner} R.~J.,  {Shabala} S.~S.,   {Krause} M. G.~H.,  2018, \mn@doi [\mnras]
  {10.1093/mnras/stx2947}, \href
  {https://ui.adsabs.harvard.edu/abs/2018MNRAS.474.3361T} {474, 3361}

\bibitem[\protect\citeauthoryear{{Werner} et~al.,}{{Werner}
  et~al.}{2010}]{werner10}
{Werner} N.,  et~al., 2010, \mn@doi [\mnras]
  {10.1111/j.1365-2966.2010.16755.x}, \href
  {https://ui.adsabs.harvard.edu/abs/2010MNRAS.407.2063W} {407, 2063}

\makeatother
\end{thebibliography}

\bsp	
\label{lastpage}
\end{document}